\documentclass[10pt,a4,twocolumn,fleqn]{article} % removed sort%compress
\usepackage[T1]{fontenc}
\usepackage[italic]{mathastext}

\usepackage[numbers,sort&compress,square]{natbib}

\usepackage[dvips]{graphicx}
\usepackage{amsmath}
\usepackage{amssymb}
\usepackage[usenames,dvipsnames]{xcolor}
\newcommand{\sj}[1]{#1}%{\textcolor{purple}{#1}}
\newcommand{\jwm}[1]{#1}%{\textcolor{purple}{#1}}
\newcommand{\ml}[1]{#1}%{\textcolor{purple}{#1}}
\newcommand{\mll}[1]{#1}%{\textcolor{purple}{#1}} % another set of ML edits

\newcommand{\sjp}[1]{#1}%\textcolor{red}{#1}}
\newcommand{\jwp}[1]{#1}%\textcolor{red}{#1}}
\newcommand{\mlp}[1]{#1}%\textcolor{red}{#1}}

\usepackage[font={small}]{caption}

\usepackage{MLHMM}
\newcommand{\SIref}[1]{\ref{#1}}
\newcommand{\SIfig}[1]{\Fig{#1}}

\usepackage{xr}
\begin{document}
% supress main text headers in the table of contents and only show SI entries
\addtocontents{toc}{\protect\setcounter{tocdepth}{-1}}

\title{Multiple LacI-mediated loops revealed by Bayesian statistics and
  tethered particle motion}

\author{\small Stephanie Johnson\footnote{Dept.~of Biochemistry and
    Molecular Biophysics, California Institute of Technology, Pasadena
    CA, USA. Present address: Dept.~of Biochemistry and Biophysics,
    University of California, San Francisco, San Francisco CA, USA.},\;
  Jan-Willem van de Meent\footnote{Dept.~of Statistics, Columbia
    University, New York, NY, USA.},\; Rob Phillips\footnote{Depts.~of
    Applied Physics and Biology, California Institute of Technology,
    Pasadena, CA, USA.},\; Chris H Wiggins\footnote{Dept. of Applied
    Physics and Applied Mathematics, Columbia University, New York,
    NY, USA},\; and Martin Lind\'en\footnote{Center for Biomembrane
    Research, Dept.~of Biochemistry and Biophysics, Stockholm
    University, Stockholm, Sweden. Present address: Dept.~of Cell and
    Molecular Biology, Uppsala University, Uppsala, Sweden. To whom
    correspondence should be addressed. Email:
    martin.linden@icm.uu.se}.\\{\small (\today)}}
% Affiliation must include:
% Department name, institution name, full road and district address,
% state, Zip or postal code, country
\date{}
\maketitle
\begin{abstract}
%!TEX root = ./HMM_NARmain.tex

%After Steph's new edits: 232 words, limit is 250
% ML 2014-01-04 : 236 words
%SJ 2014-01-07 steph counts them as 242 ... 
% ML 2014-01-27: NAR has 200 words max.

% ML 2014-02-01, after revision by SJ:

The bacterial transcription factor LacI loops DNA by binding to two
separate locations on the DNA simultaneously. Despite being one of the
best-studied model systems for transcriptional regulation, the number
and conformations of loop structures accessible to LacI remain
unclear, though the importance of multiple co-existing loops has been
implicated in interactions between LacI and other cellular regulators
of gene expression. To probe this issue, we have developed a new
analysis method for tethered particle motion, a versatile and
commonly-used {\it in~vitro} single-molecule technique. Our method,
vbTPM, \jwp{performs variational Bayesian inference in} hidden Markov
models. It learns the number of distinct states (i.e., DNA-protein
conformations) directly from tethered particle motion data with better
resolution than existing methods, while easily correcting for common
experimental artifacts.  Studying short (roughly 100 bp) LacI-mediated
loops, we \sj{provide evidence for} three distinct loop structures,
more than previously reported \jwp{in single-molecule
studies}. Moreover, our results confirm that changes in LacI
conformation and DNA binding topology both contribute to the
repertoire of LacI-mediated loops formed {\it in~vitro}, and provide
qualitatively new input for models of looping and transcriptional
regulation. We expect vbTPM to be broadly useful for probing complex
protein-nucleic acid interactions.

% version before removing abbreviations in abstract

% The bacterial transcription factor LacI loops DNA by binding to two
% separate locations on the DNA simultaneously. Despite being one of
% the best-studied model systems for transcriptional regulation, the
% number and conformations of loop structures accessible to LacI
% remain unclear, though the importance of multiple co-existing loops
% has been implicated in interactions between LacI and other cellular
% regulators of gene expression. To probe this issue, we have
% developed a new analysis method for tethered particle motion (TPM),
% a versatile and commonly-used {\it in~vitro} single-molecule
% technique. Our method, vbTPM, is based on a variational Bayes
% treatment of hidden Markov models \jwm{(HMMs)}. It learns the number
% of distinct states (i.e., DNA-protein conformations) directly from
% TPM data with better resolution than existing methods, while easily
% correcting for common experimental artifacts.  Studying short
% (roughly 100 bp) LacI-mediated loops, we \sj{provide evidence for}
% three distinct loop structures, more than previously reported at the
% single molecule level. Moreover, our results confirm that changes in
% LacI conformation and DNA binding topology both contribute to the
% repertoire of LacI-mediated loops formed in vitro, and provide
% qualitatively new input for models of looping and transcriptional
% regulation. We expect vbTPM to be broadly useful for probing complex
% protein-nucleic acid interactions.

\end{abstract}
%\section{Significance statement}
%\input{significance.tex} 
\section{Introduction}
% **************************************************************
% Keep this command to avoid text of first page running into the
% first page footnotes
%\enlargethispage{-65.1pt}
% **************************************************************
%!TEX root = HMM_NARmain.tex

Severe DNA deformations are ubiquitous in biology, with a key class of
such deformations involving the formation of DNA loops by proteins
that bind simultaneously to two distant DNA sites.  DNA looping is a
common motif in gene regulation in both prokaryotes and
eukaryotes \cite{oehler2010,schleif1992,matthews1992}. A classic
example of a gene-regulatory DNA looping protein is the Lac repressor
(LacI), which controls the expression of genes involved in lactose
metabolism
in \textit{E. coli} \cite{oehler2010,schleif1992,matthews1992}.  LacI
has two DNA binding domains, which can bind simultaneously to two
specific sites on the DNA, called operators, to form loops.  Despite
being one of the best-studied model systems of transcriptional
regulation, the mechanics of DNA looping by LacI remain incompletely
understood. One of the key outstanding issues regarding the mechanics
of loop formation by LacI is that theoretical and computational
modeling provide evidence for the existence of many conformations of
LacI-mediated loops, but it is not clear which conformations are
realized for various loop lengths, nor how many of these different
conformations are relevant for gene
regulation \textit{in~vivo} \cite{Saiz2007,zhang2006}.  Quantitative
studies of looping and transcriptional regulation would be greatly
aided by a better understanding of the \sjp{structures} of
LacI-mediated loops, as many models of looping are sensitive to
assumptions about the conformation of the protein and/or the DNA in
the loop \cite{zhang2006,swigon2006,towles2009,czapla2013}.  Moreover,
inducer molecules and architectural proteins, which are important
influencers of gene regulation \textit{in vivo}, appear to be able to
manipulate these
parameters \cite{czapla2013,Becker2005,Becker2007,Becker2008,goodson2013}.
In this work we \sj{argue} that at least three distinct loop
structures contribute to LacI-mediated looping {\it in vitro} for a
given DNA construct when the loop length is short (on the order of the
DNA persistence length), one more than the two structures that are
usually
reported \cite{wong2008,rutkauskas2009,han2009,johnson2012,johnson2013,revalee2014}.

\begin{figure}[t!]\begin{center}
  \includegraphics[width=3in]{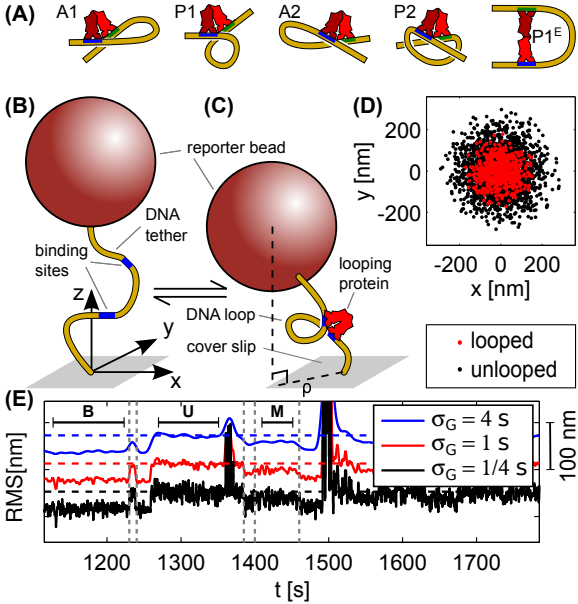}
\caption{(A) Examples of possible LacI-mediated loops, using the notation of 
Ref.~\cite{swigon2006}.  (B-C) Tethered particle motion (TPM) setup,
in which a reporter bead tethered to a cover slip by a DNA molecule is
tracked as it diffuses around the tethering point.  The formation of a
DNA loop shortens the DNA ''leash'', which narrows the distribution of
bead positions (D). The degree of restriction depends not only on the
length of the loop, but also on the relative distance and orientation
of the in- and outgoing strands, so that different loop shapes can be
distinguished. (E) Root-mean-squared (RMS) signals, time-averaged with
Gaussian filters of different kernel width $\sigma_G$ (see Methods),
for an example trace with an unlooped and two looped states (one long
stretch of each indicated by U, M, and B respectively). Horizontal
dashed lines indicate the unlooped state, offset for clarity, and
vertical ones indicate potential loop-loop interconversion
events.}\label{fig:TPMintro}
\end{center}\end{figure}

The naturally-occurring \textit{lac} operon has three operators with
different affinities for \sjp{LacI} \cite{oehler2010}, allowing
loop formation between three different pair-wise combinations \sjp{of binding sites}.  Most
studies of looping mechanics avoid this complexity by using synthetic
constructs with only two operators, but multiple loop conformations
are possible even in these simplified systems.  \jwp{The DNA-binding domains of LacI are
symmetric \cite{lewis1996}, so each operator can bind in one of two
orientations, enabling four distinct loop topologies (\Fig{fig:TPMintro}A)}. Moreover,
loops could form with the LacI protein on the inside or outside of the
DNA loop \cite{zhang2006,wong2008}.  In addition, it has been shown
that LacI has a flexible joint, allowing the V-like shape seen in the
crystal structure to adopt extended conformations as well, as in the
rightmost schematic
in \Fig{fig:TPMintro}A \cite{ruben1997,taraban2008,morgan2005,wong2008,rutkauskas2009,haeusler2012}.
Finally, the DNA binding domains seem to rotate easily in molecular
dynamics simulations \cite{villa2005}, which would help LacI to relax
strain in the DNA of the loop \cite{zhang2006,swigon2006,czapla2013}.

Different predicted loop conformations are usually classified as
differing in DNA binding topology or in LacI conformation, with a key
distinction between the two being that structures differing in DNA
topology cannot directly interconvert without LacI dissociating from
one or both operators, in contrast to those differing in LacI
conformation (e.g., V-shaped versus extended shapes), which should be
able to directly interconvert (see, for example,
Ref.~\cite{wong2008}).

The existence of multiple loop conformations for LacI-mediated
loops \textit{in vitro} has been confirmed experimentally, but
identifying these experimentally observed loops with particular
molecular structures is challenging.  One of the most widely-used
techniques for studying LacI loop conformations is a non-fluorescent
single-molecule technique called tethered particle motion
(TPM \cite{schafer1991}; see \Fig{fig:TPMintro}B), which uses the
Brownian motion of a microscopic bead tethered to the end of a linear
DNA to report on looping \cite{finzi1995}.  TPM has resolved two
looped states with a variety of synthetic and naturally-occuring DNA
sequences \cite{wong2008,rutkauskas2009,normanno2008,han2009,johnson2012,johnson2013}.
However, the structural basis of these two states is currently a
subject of debate. Importantly, direct interconversions between the
two looped states have been observed in TPM experiments with 138 bp
and 285 bp loops\mlp{. This strongly suggests that a conformational
change of LacI occurs in these loops, presumably a transition between
a V-like and a more extended state \cite{wong2008,rutkauskas2009},
since a change of loop topology would require an unlooped
intermediate.}

There is also evidence from both ensemble and single-molecule
fluorescence resonance energy transfer (FRET) experiments with
synthetic, pre-bent loop sequences, whose conformations can be
determined computationally, for at least
two \cite{edelman2003,morgan2005} and possibly
three \cite{haeusler2012} coexisting loops differing in both DNA topology
and LacI conformation.  However, it is as yet unclear which of
the structures observed by FRET correspond to the states
observed by TPM, and whether three loop conformations might also
coexist in the loops formed from generic rather than pre-bent DNA
sequences. %that have been studied by TPM.

One difficulty in determining the number of looping conformations in
TPM measurements is that not all loop conformations produce distinct
TPM signals \cite{towles2009,revalee2014}, raising the possibility
that the actual number of conformations might be greater than two.
Indeed, elastic modeling consistently predicts the coexistence of more
than two conformations for a single looping construct, either through
direct arguments (\textit{i.e.}, finding multiple loop structures with
comparable free energies \cite{czapla2013,towles2009}), or indirectly,
by predicting that the most stable V-shaped loops and the most stable
extended loops have different DNA
topologies \cite{zhang2006,swigon2006}.  In the latter case, the
lowest energy states of the V-shaped and the extended conformations
would be geometrically unable to interconvert directly with each
other, since they differ in DNA topology. Thus, previous reports of
direct loop-loop interconversions \cite{wong2008,rutkauskas2009} would
have to be explained by the existence of at least one additional loop
structure that shares a DNA topology with one of the lowest energy
states.

These considerations suggest two questions to address in order to make
progress towards identifying the loop structures relevant for
looping \textit{in vitro}: (1) is there evidence for more than two
loop structures underlying previously reported TPM data, as would be
expected from elastic modeling and from FRET results with pre-bent
sequences?, and (2) which of the observed states interconvert
directly, identifying them as differing in LacI conformation rather
than DNA binding topology?

Shorter loop lengths (\textit{i.e.}, shorter than the persistence
length of DNA, roughly 150 bp) tend to enhance the free energy
differences between loop structures, and so provide an interesting
opportunity to look for detectable signatures of additional loop
structures, and to determine which state(s) directly interconvert.  We
recently reported TPM data of two apparent looped states for loops
lengths around 100 bp \cite{johnson2012}, but the presence or absence
of direct interconversions between the two states was not
\sjp{addressed}.  Here, we revisit these data to address the
questions of direct interconversions and the number of looped states
more rigorously. We \sj{provide evidence for} the presence of a third looped state
in addition to the two previously reported, \sj{and we demonstrate} direct
interconversions between two of the three states.

\jwp{Detection of direct loop-loop interconversions requires a high
time resolution, which is especially difficult to obtain at short loop
lengths where the signal-to-noise ratio of TPM data is comparatively
small}. To meet this challenge, we have developed a powerful set of
analysis techniques for TPM data, based on \jwm{inference
in \ml{hidden Markov models (HMMs \cite{rabiner1986})} using
variational Bayesian (VB)
methods \cite{beal2003,bronson2009,okamoto2012,persson2013,vandemeent2013,vandemeent2014}.
HMMs are widely used to analyze ion channel \cite{chung_ptrslb_1990},
optical trapping \cite{smith_bpj_2001}, magnetic
tweezers \cite{kruithof_bpj_2009}, single-molecule
FRET \cite{mckinney_bpj_2006,bronson2009,okamoto2012,vandemeent2013,vandemeent2014},
and single-particle tracking \cite{persson2013} experiments.  Our
toolbox}, which we call vbTPM, offers several advantages over existing
TPM analysis techniques, including improved resolution, an objective
criterion to determine the number of (distinguishable) DNA/protein
conformational states, robustness against common experimental
artifacts, and a systematic way to pool information from many
trajectories despite considerable cross-sample heterogeneity.

vbTPM should benefit a broad community of users, as TPM is a versatile
and widely-used single molecule technique, with its simplicity,
stability, ability to measure DNA-protein interactions at very low
applied tension
\cite{segall2006,lindner2011}, and potential for \sjp{high throughput} 
\cite{plenat2012} making it an attractive tool
for \sjp{\textit{in~vitro}} studies of protein-nucleic acid
interactions that loop or otherwise deform
DNA \mlp{\cite{schafer1991,yin1994,finzi1995,vanzi2003,vanzi2006,mumm2006,broek2006,pouget2006,chu2010,rousseau2010,manzo2012,fan2012,han2009,johnson2012,johnson2013,revalee2014}}.
Moreover, our results from applying vbTPM to TPM data on short DNA
loops provide important new inputs for a comprehensive understanding
of LacI-mediated DNA looping \textit{in~vitro} and quantitative models
of transcriptional regulation \textit{in~vivo}.

% removed TPM-references
% nelson2006, brinkers2009: not a DNA-protein interaction, only TPM development
%

%Moreover, the results we describe here from the application of vbTPM
%to our TPM data with short DNA loops should provide important inputs
%to future theoretical and computational efforts towards understanding
%the mechanics of looping by LacI both \textit{in~vitro}
%and \textit{in~vivo}.

% method dev.: pouget2004 (tracking), blumberg2005 (3D), 
% RNAP: schafer1991, yin1994, tolic2003, 
% holliday junct: pouget2004
% translation by single ribosomes: vanzi2003
% restr. enzymes: broek2006 (Nael, Narl), laurens2009 (SfiI),
% laurens2012 & rusling2012 (FokI)

% lambda recombination: mumm2006, zurla2009
% lambda repressor: manzo2012
% Cre recombination:   fan2012,
% telomeres xxx: chu2010, 
% transposome: pouget2006,rousseau2010 (IS911)
%Lac:

%However, tracking in 3D \cite{lindner2013,lindner2011,blumberg2005}
%and the use nanoparticles \cite{manghi2010,lindner2011}

\section{Materials and Methods}
%!TEX root = HMM_NARmain.tex

\subsection*{TPM data}

We present new analysis of previously published data
\cite{johnson2012} for constructs that contain 100 to 109 bp of either
a synthetic random sequence called E8 \cite{cloutier2004,cloutier2005}
or a synthetic, strong nucleosome positioning sequence called 601TA
(abbreviated TA) \cite{lowary1998,cloutier2004,cloutier2005} in the
loop, flanked by the strongest naturally occurring LacI operator O1
and an even stronger synthetic operator called Oid.  We denote these
constructs E8x and TAx, where x=100-109 and refers to the length of
the loop, excluding the operators. The O1 and Oid operators are 21 and
20 bp long, so the distance between operator centers is thus x+20.5
bp. For ease of comparison between our results and others', we use
loop length, not distance between operator centers, when quoting
other's results. The \textit{in vitro} affinities of LacI for the O1
and Oid operators are roughly 40 and 10 pM respectively
\cite{johnson2012,whitson1986a,whitson1986b,hsieh1987,frank1997}.  The
total lengths of the DNA tethers range from 458-467 bp, depending on
the length of the loop \cite{johnson2012}.

For every tethered DNA, we collected 10 minutes of calibration data in
the absence of LacI, followed by roughly 20 to 100 minutes of looping
data in the presence of 100~pM LacI, purified in-house.  Data sets for
each loop length typically contain 50-100 TPM trajectories.  We used a
standard brightfield microscopy-based TPM setup, where 490~nm diameter
polystyrene beads are tracked in the xy-plane with video microscopy at
30~Hz, and the resulting trajectories then drift-corrected using a
first-order Butterworth filter with a 0.05~Hz cutoff frequency (see
Ref.~\cite{johnson2012} for detailed experimental and analysis
procedures).  As noted below, this drift-corrected data was used as
the input for the HMM analysis (and not the subsequently
Gaussian-filtered RMS trajectories that are described in
Ref.~\cite{johnson2012}).

In addition to the pre-existing data, we also obtained calibration
trajectories from constructs with total lengths \sjp{450~bp} (``E894'' of
Ref.~\cite{johnson2012}), 735~bp (``wild-type'' of
Ref.~\cite{johnsonPhD}), and 901~bp (``PUC306'' of
Refs.~\cite{han2009,johnsonPhD})). Data for these constructs were
obtained in the absence of LacI only.

% OidE8LO1 bp count:
% 144 (flank 1) + 20 (Oid) + L + 21 (O1) + 172 (flank 2) =358+L

\subsection*{RMS analysis} 
 The root-mean-square (RMS) trace of a tether is the square root of a
 running average of the variance of the bead's position,
 $\sqrt{\mean{\rho^2}}$. We followed the procedures of
 Ref.~\cite{johnson2012}, \sj{in which $\rho$ was calculated from drift-corrected $x$ and $y$ bead positions, as described in the previous section, and then convolved with a Gaussian filter,} except \sj{here} we varied the standard deviation
 $\sigma_G$ of the Gaussian filter kernel for the running average\sj{, rather than keeping it fixed at 4~s as in \cite{johnson2012}}. To
 count the number of states, we determine the number of peaks in RMS
 histograms by eye.

\subsection*{Diffusive HMM for single trajectories}
\jwm{vbTPM uses a \jwp{diffusive} HMM to describe the bead motion and looping
  kinetics in a manner that directly models bead positions instead of
  RMS traces. In an HMM, kinetics are modeled} by a discrete Markov
process $s_t$\ml{, $t=1,2,\ldots,T$,} with $N$ states (\textit{e.g.},
$s_t=1$ when unlooped, $s_t=2$ when looped, \textit{etc.}), a
transition probability matrix $\matris{A}$, and \jwp{an} initial state
distribution $\vec{\pi}$,
\begin{equation}\label{eq:eoms}
  p(s_t|s_{t-1},\matris{A})=A_{s_{t-1}s_t},\quad p(s_1|\vec\pi)=\pi_{s_1}.
\end{equation}
\jwm{The physics specific to TPM are contained in} the emission model,
which describes the motion of the bead for each hidden state. We use a
discrete-time model of over-damped 2D diffusion in a harmonic
potential that has been suggested as a simplified model for TPM
\cite{beausang2007b,lindner2013}ml{. This means that the probability
  distribution of each bead position is Gaussian, and depends
  conditionally on the hidden state and previous position,} 
\begin{equation}\label{eq:eomx}
 p(\x_t|\x_{t-1},s_t,\vec{K},\vec{B})
 =\frac{B_{s_t}}{\pi}
 e^{-B_{s_t}(\x_t-K_{s_t}\x_{t-1})^2}.
\end{equation}
%,
%\begin{equation}
%  \x_t=K_{s_t}\x_{t-1}+\vec{w}_t/(2B_{s_t})^{1/2},
%\end{equation}
%where the index $s_t$ indicate parameters that depend on the hidden
%state, and $\vec{w}_t$ are uncorrelated Gaussian random vectors with
%unit variance. Hence, the probability distribution of each bead
%position conditional on the hidden state and previous position is
%given by
%\begin{equation}\label{eq:eomx}
% p(\x_t|\x_{t-1},s_t,\vec{K},\vec{B})
% =\frac{B_{s_t}}{\pi}
% e^{-B_{s_t}(\x_t-K_{s_t}\x_{t-1})^2}.
%\end{equation}
The emission parameters $K_j$ and $B_j$ are related to the spring and
diffusion constants \ml{of the corresponding hidden states}. More
insight into their physical meaning can be gained by noting that with
a single hidden state, Eq.~\ref{eq:eomx} describes a Gaussian process
with zero mean and
\begin{align}
&\mathrm{RMS}=
\sqrt{\mean{\rho^2}}=\sqrt{\mean{\x^2}}=(B(1-K^2))^{-1/2},\nonumber\\
&\mlp{\mean{\x_{t+m}\cdot\x_t}/\langle \x^2\rangle
=K^{|m|}\equiv e^{-|m|\Delta t/\tau}},\label{eq:1state}
\end{align}
where $\Delta t$ is the sampling time, and \mbox{$\tau=-\frac{\Delta
    t}{\ln K}$} is a bead correlation time \mlp{(see
  Sec.~\ref{sec:RMSderivation} \sjp{in the Supporting Information (SI)})}.  This model captures the diffusive
character of the bead motion while retaining enough simplicity to
allow efficient statistical analysis.

\subsection*{\jwp{Inference and model selection}}
To analyze TPM trajectories using the above model, we apply a VB
technique \cite{beal2003} that has previously been used in the
analysis of other single-molecule data
\cite{bronson2009,okamoto2012,persson2013,vandemeent2013,vandemeent2014},
but has not been applied to \jwp{TPM} data so far.  VB methods can
determine both the most likely number of hidden states $N$ and the
most likely parameters $\theta = \{\matris{A}, \vec{\pi}, \vec{K},
\vec{B}\}$ for the model.  Models with more states and parameters can
generally model the data more closely, but may overfit the data by
attributing noise fluctuations to separate states.  VB methods perform
model selection by ranking models according to a lower bound $F_N$
on the log evidence $\ln L_N$.  The evidence $L_N$ is the marginal
probability of observing the measurement data, obtained by integrating
out all model parameters $\theta$ and hidden state sequences $\{s_t\}$
from the joint probability $p(\{\x_t\}, \{s_t\}, \theta \,|\, N)$,
\begin{equation}\label{eq:F}
  F_N \lesssim \ln L_N=\ln \sum_{s_1,s_2,\ldots}\int
  p(\{\x_t\},\{s_t\}|\theta,N)p(\theta|N) d\theta .
\end{equation}
The model with the highest lower bound log evidence $F_N$ can be
interpreted as the model that exhibits the best ``average'' agreement
with the data over a range of parameters, thereby eliminating models
that overfit the data and only show good agreement for a narrow
parameter range.  VB analysis requires us to parameterize our prior
knowledge (or ignorance) about parameter values in terms of prior
distributions $p(\theta \,|\, N)$. We choose ``uninformative'' priors to
minimize statistical bias.  VB analysis also yields parameter
information in terms of (approximate) posterior distributions on
$\theta$, which are optimized numerically to maximize $F_N$ when
fitting a model to data. We generally report parameter values as
expectation values of these distributions. Further details are given
in the \sjp{SI} and software documentation (see below).

\subsection*{Downsampling}
To decrease the computational cost associated with analysis of large
data sets, we downsample by restricting the hidden state changes to
occur on multiples of $n$ data points. By downsampling only the hidden
states, and not the TPM data, we avoid discarding valuable information
about bead relaxation dynamics \cite{lindner2013,beausang2007}. We use
$n=3$ except where noted otherwise.  With an original sampling
frequency of 30 Hz and $K\gtrsim 0.4$ ($\tau\gtrsim 1/30$ s) in our
data (see Results), the shortest possible state lifetime (1/10 s after
downsampling) is thus at most three times larger than the bead
correlation time.

\subsection*{Synthetic data}
We generate synthetic data by direct simulation of
Eqs.~\eqref{eq:eoms} and \eqref{eq:eomx}, followed by application of a
first-order Butterworth filter with 0.05 Hz cutoff frequency to
simulate drift-correction \cite{han2009,johnson2012}.  To generate
reasonable parameter pairs, we use the empirical fit
$\tau=0.018~\textrm{ RMS}-0.079$, with $\tau$ in seconds and RMS in
nm, and then compute $K,B$ from \Eq{eq:1state}. For analysis, we use
the same settings (priors, \textit{etc.}) as for real data.

\subsection*{\jwp{Pooled analysis of multiple trajectories}}
To \jwp{make full use of} the high-throughput capabilities of TPM, it is
advantageous to pool information from many trajectories in a
systematic way. Indeed, we will see below that this \jwp{is} necessary to
unambiguously resolve direct interconversions between looped
states. \jwp{Two problems must be solved in order to pool information 
from multiple trajectories.} First, TPM data contain artifacts, \textit{e.g.} transient
sticking events or tracking errors (described in more detail
below). Such spurious events are specific to each trajectory, and
should not be pooled.  Second, variations in bead size, attachment
chemistry, \textit{etc.},~create significant variability between beads in
nominally equal conditions (\textit{e.g.} DNA construct length and LacI
concentration \cite{johnson2012}), \sjp{making it in}feasible to
fit a single model to multiple trajectories even without spurious
events.

To address the first problem, we extend the single trajectory HMM with
a second type of hidden state, $c_t$, such that $c_t=1$ indicates
genuine looping dynamics governed by the simple model described
above. When $c_t>1$, the bead motion is instead assumed to arise from
some kind of measurement artifact, which is modeled by a different set
of emission parameters $\hat B_{c_t},\hat K_{c_t}$.  We assume the
genuine states, $s_t$, to evolve independently of spurious
events. Similarly, spurious events $c_t>1$ can interconvert
independently of the underlying genuine state, but transitions out of
$c_t=1$ depend on $s_t$, to allow for possibilities such as transient
sticking events being more frequent in a looped state when the bead is
on average closer to the cover slip. \ml{ These assumptions mean that
  the joint transition probability of $s_t,c_t$ factorizes as}
\begin{equation}\label{eq:factorial1}
  p(s_{t+1},c_{t+1}|s_t,c_t)=p(s_{t+1}|s_t)p(c_{t+1}|s_t,c_t).
\end{equation}
\ml{We therefore refer to it as a (variant of a) factorial HMM
  \cite{ghahramani1997}.}  As before, $p(s_{t+1}|s_t)=A_{s_ts_{t+1}}$,
but transitions involving the spurious states are described by two new
transition matrices $\hat{\matris{A}}$ and $\hat{\matris{R}}$,
%with $p(s_{t+1}|s_t)=A_{s_ts_{t+1}}$ as earlier, and
\begin{equation}\label{eq:factorial2}
  p(c_{t+1}|s_t,c_t)
  =\left\{
  \begin{array}{ll}
    \hat A_{s_tc_{t+1}},&\text{ if $c_t=1$,}\\
    \hat R_{c_tc_{t+1}},&\text{ if $c_t>1$,}\\
  \end{array}
  \right.
\end{equation}

To deal with bead-to-bead variability, we adopt an empirical Bayes
(EB) approach that derives from a recently-developed analysis
technique for single-molecule FRET data
\cite{vandemeent2013,vandemeent2014}.  In EB analysis, the prior is
interpreted as the distribution of model parameters across the set of
trajectories, and is learned from the data to maximize the total lower
bound log evidence.  In this manner, similarities between trajectories
are exploited to obtain more accurate parameter estimates.  We
restrict EB analysis to transition probabilities and emission
parameters of the genuine states \ml{($s_t$ in
  Eqs.~\ref{eq:factorial1}-\ref{eq:factorial2})}, while priors
describing spurious states are held fixed.

Pooled analysis using EB and the factorial model is performed in
\ml{four steps, summarized in Sec.~\SIref{sisec:workflow}}. First, we
perform VB analysis, learning the optimal number of states for each
trajectory. Second, looped states and artifact states are classified
using an automated procedure (see \Eq{eq:genuinecriterium} below), and
verified manually using a graphical tool.  In practice, very few
corrections to the automated classification are needed. \ml{Third,
  factorial models are generated by translating the spurious states of
  the simple HMMs into $c_t>1$-states
  (Eqs.~\ref{eq:factorial1}-\ref{eq:factorial2}), and reconverged
  using a VB algorithm. Finally, these factorial models are used as an
  initial guess for the EB algorithm. Since EB analysis requires all
  models to have the same number of genuine states, some factorial
  models also have to be extended with extra unoccupied states.}
Information can then be extracted from the optimized prior
distributions. Further details are given in the software
documentation.

\subsection*{Implementation} 
vbTPM runs on Matlab with inner loops written in C, and includes a
graphical tool for manual state classification. Source code and
software documentation are available at vbtpm.sourceforge.net.

\section{Results}
%!TEX root = HMM_NARmain.tex

% confirmed RMS users:
% \cite{towles2009,wong2008,rutkauskas2009,normanno2008,han2009,johnson2012,manghi2010,vanzi2006,broek2006},

\subsection*{Improved resolution on synthetic data}
\addcontentsline{toc}{subsection}{Improved resolution on synthetic data}

A simple and common way to analyze TPM data is in terms of RMS values,
which are the square root of the bead position variance, or the
projected distance $\rho$ between the bead center and tether point
(\Fig{fig:TPMintro}E). Transitions can be extracted by thresholding
RMS traces, and the number of states by counting peaks in RMS
histograms
\cite{finzi1995,wong2008,vanzi2006,laurens2009,laurens2012,johnson2012,johnson2013}.
However, the RMS signal must be smoothed in order for the transitions
to appear. This degrades the time resolution \cite{manghi2010}\ml{,
  and a direct analysis of bead position traces, such as vbTPM, would
  likely do better in this respect \cite{manzo2010}.} As noted above,
  this is of particular interest when determining whether or not
  apparent loop-loop interconversions are in fact separated by short
  unlooped intermediates.

We have tested vbTPM on synthetic data, and compared its ability to
resolve close-lying states with that of the RMS histogram method. Two
states can be difficult to resolve either due to similar RMS values or
short lifetimes. Our state detection tests (see
\SIfig{sifig:resolution}-\SIref{sifig:HMMresolution}) show that vbTPM
offers a great improvement over RMS histograms in the latter case,
which is precisely the case that matters most for the question of
direct interconversions that we address here. For example, two states
separated by 40~nm are resolved by vbTPM at a mean lifetime of about
0.5~s, while lifetimes of 4-8~s are necessary for states to be
resolvable in RMS histograms (\SIfig{sifig:resolution}). This order of
magnitude improvement mainly reflects the detrimental effects of the
low-pass filter used in the RMS analysis \sj{(see RMS analysis in
  Materials and Methods)}. The difference diminishes for more
long-lived states, and with a mean lifetime of 30~s, the spatial
resolution is about 15~nm for both methods
(Fig.~\SIref{sifig:RMSresolve} and~\SIref{sifig:dRMS40rms}).

Our tests with synthetic data further show that the parameters,
including transition rates, are faithfully recovered by vbTPM, and
that all of these results are insensitive to downsampling by the
factor of three that we use when analyzing real data
(\SIfig{sifig:dRMS40rms}-\SIref{sifig:dRMS40Aij}).

\subsection*{Detection of experimental artifacts}
\addcontentsline{toc}{subsection}{Detection of experimental artifacts}
A striking illustration of the improved time resolution of vbTPM is
the ability to detect and classify short-lived experimental artifacts
in the data.  Our normal TPM protocol starts with a short calibration
run in the absence of the looping protein for quality control reasons
\cite{johnson2012}. Here, we expect only one state, that of the fully
extended tether. However, analyzing calibration data for three
different construct lengths, we find more than one state in most
trajectories, although a single state usually accounts for most
($\sim$99\%) of the trajectory.

\begin{figure}[t!]
\begin{center}
\includegraphics[width=3in]{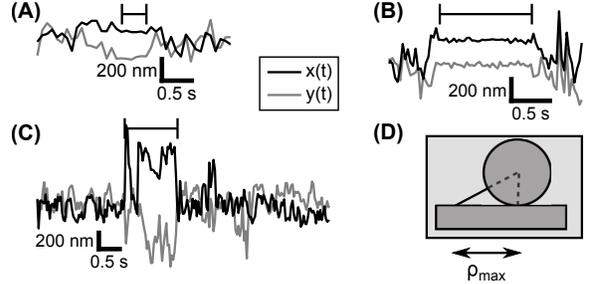}
\caption{Examples of spurious events in calibration data
  (\textit{i.e.}, in the absence of repressor).  Spurious events are
  marked by horizontal black lines above the blue and red time traces
  of the bead's $x$ and $y$ positions.  (A,B) show ``sticking events''
  (non-specific, transient attachments of the bead to the surface, the
  DNA to the bead, \textit{etc}), while (C) contains an excursion
  larger than the physically possible maximum, $\rho_\text{max}$, as
  shown in (D). This could be due to a tracking error, for example
  when an untethered bead diffuses through the field of view. \ml{Note
    that the events shown here are all on the second time scale, and
    hence undetectable with the temporal resolution of about 11 s in
    our previous RMS-based analysis
    \cite{johnson2012}.}}\label{fig:spuriousevents}
\end{center}
\end{figure}

\begin{figure}[b!] 
  \begin{center}
    \includegraphics[width=3in]{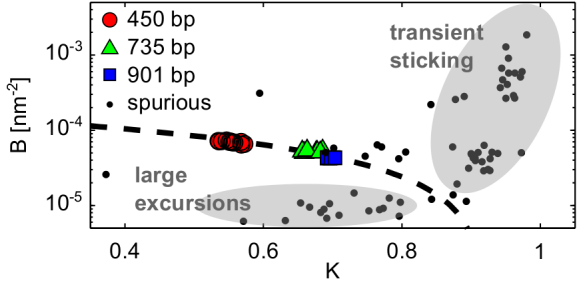}
  \end{center}
\caption{Scatter plot in the $(K,B)$ plane of genuine and spurious
  states in trajectories without LacI from three different tether
  lengths. The genuine states, colored according to tether length, are
  defined as the most long-lived state in each trajectory, and fall
  close to the empirical fit $B=(1.84-2K)\times 10^{-4}$ nm$^{-2}$
  (dashed line, note log-scale on the B-axis). Spurious states (dots)
  scatter off of this line. Gray ellipsoids indicate rough parameter
  trends for sticking and tracking errors (large excursions)
  respectively.}\label{fig:HMMemission}
\end{figure}

Inspection of the coordinate traces (that is, the $x$ and $y$
positions of the bead as functions of time) reveals the dominant state
to correspond to normal, ``genuine'' bead motion, while the extra
``spurious'' states are associated with obvious irregularities in the
data. Many of these are too short-lived to show up in the
time-averaged RMS traces.  Almost all can be interpreted as either
transient sticking events (\Fig{fig:spuriousevents}A-B), where the
motion in $x$ and $y$ simultaneously and abruptly goes down
dramatically, or brief excursions beyond the limit set by the tether
length (\Fig{fig:spuriousevents}C-D), caused by breakdowns of the
tracking algorithm when\sjp{, for example,} free beads diffuse through
the field of view.  Some spurious events are described as more than
one state in the vbTPM analysis. A scatter plot of the emission
parameters $K$ and $B$ for detected states (see
\mlp{Eqs.~(\ref{eq:eomx},\ref{eq:1state}}) show different patterns for
genuine and spurious states (\Fig{fig:HMMemission}). Genuine states
fall along a curve in the $K,B$ plane, while the spurious states
scatter. This makes physical sense, since the genuine dynamics are
governed by a single parameter, the effective tether length, while the
spurious states are of diverse origins.  This pattern persists also in
trajectories with looping, with the genuine looped states continuing
along the curve indicated by the calibration states
(\Fig{fig:lacstates}A).

The $K,B$ values of different trajectories vary significantly, but it
turns out that within fitting uncertainty, most states of individual
trajectories satisfy
\begin{equation}\label{eq:genuinecriterium}
K_{\ml{gen.}}\le K_{cal.}\text{, and }B_{\ml{gen.}}\ge B_{cal.},
\end{equation}
\ml{with $(\cdot)_{cal.}$ and $(\cdot)_{gen.}$ denoting genuine state
  parameters of calibration and looping trajectories,
  respectively. Most} spurious states violate at least \sj{one} of
these inequalities. An intuitive rationale for this rule is that $K$
($B$) tends to decrease (increase) with decreasing tether length as
seen in \Fig{fig:HMMemission}. Looping decreases the effective tether
length, as does the slight bending of the operator sites upon LacI
binding \cite{lewis1996,johnson2012}.

The upshot of the different behaviors of genuine and spurious states
shown in Figs.~\ref{fig:HMMemission} and~\ref{fig:lacstates}(A) is
that \Eq{eq:genuinecriterium}, plus an additional lower threshold on
RMS values (see \Eq{eq:1state}) to catch sticking events near the
tethering point, can be used to computationally label genuine versus
spurious states. Very few exceptions remain to be corrected manually.
\ml{While spurious states make the HMM analysis more complicated, they
  constitute a sufficiently minor fraction of most trajectories, such
  that their presence does not significantly affect the average
  looping properties (see
  \SIfig{sifig:spuriousstats}-\ref{sifig:spuriousdiff}), and hence
  their presence does not invalidate previous TPM results that did not
  remove them.}

%\sj{Although as shown in Fig.~\ref{??} in the SI these
%    spurious states constitute a relatively minor fraction of the
%    total time of a given trajectory, and so their detection and
%    removal in our vbTPM analysis does not invalidate previous TPM
%    results that did not remove them, they do impede further HMM
%    analysis, and particularly the combined analysis of multiple
%    trajectories at once, described below. }
  
\subsection*{More than two looped states}
\addcontentsline{toc}{subsection}{More than two looped states} 
%\jwp{our vbTPM} technique 
We \mlp{used vbTPM} to examine looping at 100~pM LacI in E8x and TAx
constructs, where ``x'' indicates the loop length, ranging from x=100
to 109 bp \cite{johnson2012}, and E8 and TA are two different DNA
sequences in the loop (see Materials and Methods). We \jwp{applied}
\Eq{eq:genuinecriterium} complemented by visual inspection to identify
genuine states, and from now on, we will understand all ``states'' to
be genuine unless stated otherwise. Most trajectories exhibit one to
three states in the presence of LacI.

We discard trajectories with only one state, as a complete lack of
looping activity might reflect defective constructs, surface
attachment, or LacI molecules \cite{johnson2012}. We also discard a
small number of trajectories with four states, where inspection
reveals either a state split by bursts of spurious events (resulting
in artificial differences in state lifetimes), or a genuine-looking
state with very low RMS that can be attributed to a sticking event
near the tethering point.  Thus, our HMM analysis is at first glance
consistent with earlier findings of two distinguishable looped states
in these constructs \cite{johnson2012}. We denote the states from
trajectories with three states unlooped (U), ``middle'' (M), and
``bottom'' (B), in keeping with the conventions of
\cite{johnson2012,johnson2013}, in which ``middle'' and ``bottom''
refer to the tether lengths of the two distinguishable looped states
relative to the unlooped state.

\begin{figure}[t!]
\includegraphics{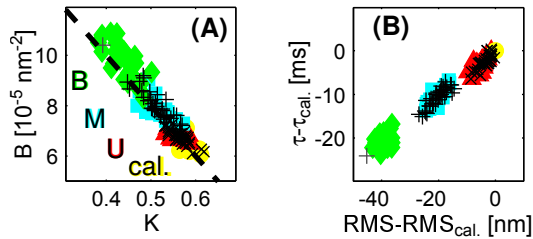}
\caption{Clustering of LacI-induced looped and unlooped genuine states
  in the E8106 construct. States U, M, and B in three-state
  trajectories are represented as filled symbols, while states in
  two-state trajectories are plotted as +'s (for the looped state) and
  x's (for the unlooped state).  (A) Raw emission parameters
  $K,B$. The dashed line is the linear fit from
  \Fig{fig:HMMemission}. (B) Same states as in A, but plotted as RMS
  values and relaxation times $\tau$ (see \Eq{eq:1state}) relative to
  the calibration (that is, no-LacI) states for each tether. From now
  on, we will plot states in these more intuitive and homogeneous
  terms.}\label{fig:lacstates}
\end{figure}

We find, however, that not all \sjp{of the remaining} trajectories in a population show
all three states; some have only one of the two looped states.
The two- \sjp{versus} three-state-containing
trajectories display a striking pattern that we will introduce using
the E8106 construct. As shown in \Fig{fig:lacstates}A, a scatter plot
of the emission parameters for three-state trajectories \ml{(colored
  symbols)} produces partly overlapping clusters in the
$K,B$-plane\ml{,} corresponding to the three observed states (U, M,
B). Some contributions to the parameter noise, such as bead size
variations, might be correlated between states, and can thus \sj{be}
reduced by normalization.  Indeed, visualizing the states relative to
their calibration states (\Fig{fig:lacstates}B) produces
well-separated state clusters.  These clusters allow us to classify
the states in the trajectories with only two states \ml{(+ and x in
  \Fig{fig:lacstates})}, by comparison to the clusters formed by the
three-state trajectories. In 37 out of 38 two-state trajectories, the
two states coincide with the U and M states.  That is, in trajectories
that only exhibit one of the two looped states, for the E8106
construct \sjp{that} looped state  is {\it always} the ``middle''
state.

\sj{One possible explanation for this pattern is that it results from
  insufficiently equilibrated three-state kinetics---that is, all
  two-state trajectories are really three-state trajectories that were
  not observed long enough. In Sec.~\SIref{sitext:equilibration}, we
  show using simulated data that under this null hypothesis we would
  expect significantly more three-state trajectories than we actually
  observe in most constructs. In other words, the number of 2-state
  trajectories found in our analysis is not consistent with a simple
  equilibration effect. We hypothesize instead that there are two
  underlying populations in our data, one population that has two
  states (one looped and one unlooped), and one population with three
  states.}

Similarly, \sjp{we find that} a sub-population of LacI \sjp{that} is
somehow unable to support the B state is \sjp{also} unlikely, as
different cluster patterns appear with other \sjp{loop lengths and}
sequences. As shown in \Fig{fig:E8series}, when we subject E8 and TA
constructs spanning one helical repeat to the same analysis, we see
some constructs (e.g., E8103, TA104, E8105, TA106) mimic the 2+3-state
pattern of E8106, but in others (E8100-101, TA100-101, TA109) the
looped state in two-state trajectories is the B rather than M
state. Moreover, while there is also one case for each sequence with
almost exclusively 3-state (E8107) or 2-state (TA105) trajectories,
the identity of the looped state in two-state trajectories exhibits a
clear phasing that correlates with loop length, and therefore with the
helical repeat of the DNA. In particular, when the operators are
in-phase and looping is maximal, demonstrated in our previous work to
occur around \sjp{106~bp} \cite{johnson2012}, the looped state in
two-state trajectories is predominately the M state, whereas when the
operators are out-of-phase, around 100 or 110 bp \cite{johnson2012},
two-state trajectories contain primarily the B state as the looped
state.

We propose a structural explanation for \mll{these observations},
namely, that the M-state in trajectories exhibiting only two states
corresponds to a different loop structure than the M-state in
trajectories with three states\ml{, and that interconversion between
  the two- and three-state regimes occurs slowly, via multiple
  unlooped states, as sketched in \Fig{fig:23pattern}}.  A further
line of evidence supporting this explanation concerns the question of
whether or not the M and B states in three-state trajectories
interconvert: if the M state can interconvert with the B state in
three-state trajectories, but the M state in two-state trajectories
never interconverts with the B state (because these trajectories show
no B state), then it is likely that these two M states
(interconverting and not interconverting) are structurally different.
Moreover, as noted in the Introduction, the question of direct
interconversions can provide insight into what structures might
underlie the interconverting and non-interconverting M and B states:
if two looped states interconvert without passing through the unlooped
state, this would indicate that the involved states have the same DNA
binding topologies, since a change of binding direction would require
an unlooped intermediate.  To address these questions, we now ask if
the looped states in three-state trajectories interconvert
directly\sj{---that is, if one of the blue states in
  \Fig{fig:E8series} can be followed by a green state without passing
  through a red state, and similarly for green to blue}.

\subsection*{Direct loop-loop interconversions}
\addcontentsline{toc}{subsection}{Direct loop-loop interconversions}
Detecting direct interconversions between looped states is
difficult. Potential events can be spotted in RMS traces, but as
illustrated in \Fig{fig:TPMintro}E, their interpretation depends on
the filter width $\sigma_G$, and we cannot exclude the presence of
short unlooped intermediates by eye.  To test whether the increased
temporal resolution of our \jwp{HMM-based analysis} could improve upon the
detection of short unlooped intermediates, we generated synthetic data
using realistic parameters obtained from %single-trajectory 
\sjp{the} E8106 and E8107 constructs with three genuine states, with spurious
states removed. The transition probabilities $A_{ij}$ from these fits
allow loop-loop interconversions, typically no more than ten per
trajectory, but we also generated data without interconversions by
setting $A_{BM}=A_{MB}=0$.

\begin{figure}[b!]
  \begin{center}
  \includegraphics[width=3in]{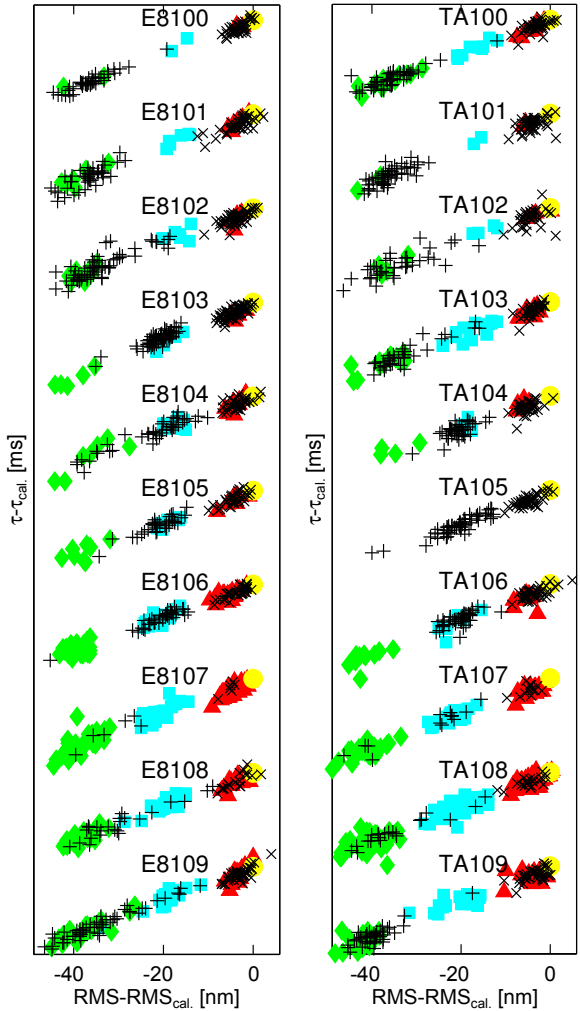}
  \caption{Clustering of looped and unlooped states for E8x and
    TAx constructs, with loop lengths x=100-109 bp. The states
    are colored and aligned as in \Fig{fig:lacstates}B, and offset
    in the $\tau$ direction for clarity.}\label{fig:E8series}
    \end{center}
\end{figure}

\begin{figure*}[t!]
\begin{minipage}[c]{\columnwidth}
\includegraphics[width=\columnwidth]{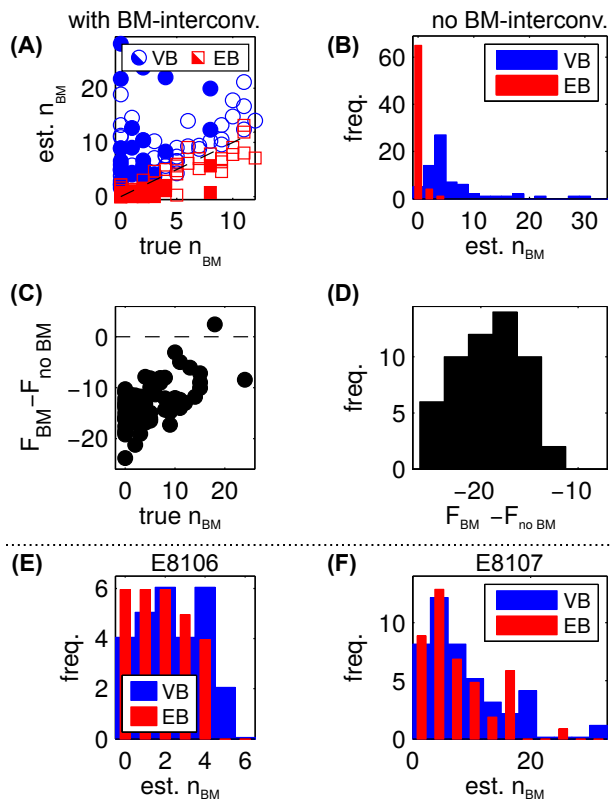}
\end{minipage}\hfill
\begin{minipage}[c]{\columnwidth}
\caption{Detecting direct loop-loop interconversions. (A,B) Counting
  the number of $B \leftrightharpoons M$ interconversions, $n_{BM}$,
  detected in synthetic data, with (A) and without (B) such
  transitions \sj{actually} present, when trajectories are considered
  analyze all trajectories from the same data set at once
  (``EB''). The dashed black line in (A) indicates where the estimated
  number of interconversions equals the true number. \sj{Since most
    blue points lie above this line, the VB approach overestimates the
    number of true interconversions; but the EB analysis either
    accurately counts such transitions, or slightly underestimates
    them.} \ml{Filled and open symbols in (A) refer to trajectories
    created from E8106 and E8107 trajectories respectively.} (B) shows
  a histogram \sj{of the number of direct interconversions per
    trajectory} rather than a scatter plot, because the true number of
  interconversions is zero\sj{; here the EB analysis accurately
    estimates that there are few or no interconversions, whereas the
    VB approach incorrectly assumes direct interconversions where
    there are in fact none}. (C,D) VB analysis of single synthetic
  trajectories prefers models without \sjp{BM-i}nterconversions, whether
  they are present (C) or not (D), probably since they are rare
  events. Every point and histogram count represent\sjp{s} a single
  trajectory, and $F_{(\ldots)}$ is the approximate log evidence,
  \Eq{eq:F}, for the different models.  Higher $F$-values indicate
  better fits, so $F_{BM}<F_{no\;MB}$ means that models with no
  interconversions are preferred by this analysis. (E,F) Analysis of
  real data yields a substantial number of interconversions even with
  the EB scheme, a strong indication that they are in fact present.
  \sj{That is, histograms of the number of direct interconversions per
    trajectory have significant weight at values above the zero bin,
    when estimated both by the VB approach (which tends to
    overestimate interconversions) and the EB approach (which
    accurately or slightly underestimates them).}}\label{fig:nMB}
\end{minipage}\hfill
\end{figure*}

Refitting these synthetic data sets with our standard settings, we
find that the HMM algorithm over-counts the number of looped state
interconversions, $n_{BM}$, even when they are absent in the data
(\Fig{fig:nMB}A-B).  Moreover, models that disallow direct
BM-interconversions generally get higher F-values (related to goodness
of fit; see Eq.~\eqref{eq:F}) than models that allow \sjp{interconversions,}
%when fitted to single trajectories, 
\sjp{even} when such interconversions
are actually present (\Fig{fig:nMB}C-D).  Thus, we cannot settle the
question of direct loop-loop interconversions by analysis of single
trajectories, probably because the number of such interconversions per
trajectory are too few in our data and in the synthetic data we create
from it.

  \begin{figure}[t!]
\begin{center}
\includegraphics[width=3in]{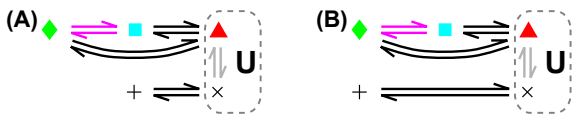}
\end{center}
\caption{\sj{Proposed kinetic models for the ``2+3'' pattern of states
    observed in \Fig{fig:E8series}, with slow interconversions (gray
    arrows) between two- and three-state trajectories occurring via
    multiple unlooped states.  Symbols and colors follow those of
    \Fig{fig:E8series}. (A) Kinetic model for in-phase operators, {\it
      e.g.} around 106~bp loops, where looping is maximal and the
    looped state in two-state trajectories is the M state, represented
    by a ``+'' as in \Fig{fig:E8series}. (B) Kinetic model for
    out-of-phase operators, e.g., around 100 or 110 bp loops, where
    looping is minimal and the looped state in two-state trajectories
    is the B state, again represented by a ``+''.  Purple arrows
    represent putative direct loop-loop interconversions, whose
    existence is explored in the last section of the
    Results.}}\label{fig:23pattern}
\end{figure}

To \jwp{overcome these limitations, we \mlp{perform pooled} analysis
  of multiple trajectories.}  The difficulty in this analysis is that
we cannot simply fit a single model to multiple trajectories, because
of the large bead-to-bead variations in motion parameters ($K$, $B$)
seen in \Fig{fig:lacstates}A, and the varying numbers of spurious
states in different trajectories \sj{seen in \Fig{fig:HMMemission},
  which differ in both number and parameter values for each
  trajectory}.  To solve these problems, we first extend our HMM to
split spurious and genuine states into two separate hidden processes
(\ml{what we call a factorial HMM;} see Methods).  Second, we
implement an empirical Bayes (EB) approach
\cite{vandemeent2013,vandemeent2014} (see Methods), which optimizes
the prior distributions based on the variability of genuine states in
different trajectories. This allows information from the whole data
set to be used in interpreting each single trajectory, and has been
shown to greatly improve the resolution in single molecule FRET data
\cite{vandemeent2013}.

Analysis of synthetic data, where the true number of interconversion
events is known, shows clear improvements when using \sjp{our} EB
analysis in comparison to normal VB methods that analyze each
trajectory individually. As shown in \Fig{fig:nMB}A, the tendency to
over-estimate the number of BM-interconversions is eliminated when the
EB scheme is applied, and almost no such transitions are detected in
trajectories where they are absent (\Fig{fig:nMB}B).  This shows that
the EB scheme can reliably detect the presence of direct
\sjp{BM-i}nterconversions\ml{, although it tends to undercount when
  transitions are very rare (see also
  \SIfig{sifig:nMBE8107doublerate}).}

\jwp{EB analysis of experimental data shows} a substantial number of
direct BM-interconversions in \mll{three-state trajectories from E8106
  and E8107 (\Fig{fig:nMB}E-F), as well as from the other constructs
  where there are a significant number of three-state trajectories
  present
  (\SIfig{sifig:E8interconversions}-\SIref{sifig:TAinterconversions}). This
  is a strong indication that direct loop-loop interconversions
  \sjp{do} occur in the short\sjp{-}loop\sjp{-}length regime studied
  here.}

%Hence, we conclude 
\mll{This evidence \sjp{for direct} loop-loop interconversions,
  \sjp{taken together with} the overrepresentation of two-state
  trajectories discussed in the previous section, lead us to
  hypothesize that} most constructs in \Fig{fig:E8series} exhibit at
least three distinct loop structures, one more than previously
reported in a single construct by TPM
\cite{han2009,johnson2012,rutkauskas2009,wong2008,johnson2013,revalee2014}: an
M and a B state that can interconvert without an unlooped
intermediate, suggesting that they share the same DNA topology but
different LacI conformations ({\it e.g.} a V-shaped and an extended
conformation); and an M (for in-phase operators) or B (for
out-of-phase operators) state that cannot directly interconvert with
another looped state.

\section{Discussion}
%!TEX root = HMM_NARmain.tex

We have developed a Bayesian analysis method for TPM data based on
hidden Markov models\ml{,} called vbTPM. A major advance offered by
our method is improved time resolution, which stems from our direct
analysis of position data, thus avoiding the time-averaging required
to produce readable RMS traces (\Fig{fig:TPMintro}). We are not the
first to exploit this possibility.  Beausang and
Nelson \cite{beausang2007} used manually curated training data to
construct detailed models of the diffusive bead motion for the looped
and unlooped states, and combined them with a two-state HMM to extract
interconversion rates.  Manzo and Finzi \cite{manzo2010} modeled bead
positions as uncorrelated zero-mean random variables, and used
change-point and hierarchical clustering methods to segment TPM
position traces in order to extract \sjp{dwell time statistics.} %for further analysis.

Our new analysis tool improves on previous methods in several
ways. Compared to the change-point method \cite{manzo2010}, we use a
noise model that accounts for correlations in the bead motion, which
eliminates the need to filter out short dwell times. Compared to the
previous HMM treatment \cite{beausang2007}, which used a more detailed
dynamical model, vbTPM does not require curated training
data. Instead, it learns the number of states directly from the data
along with all other model parameters in a statistically principled
way, using a variational Bayes treatment of
HMMs \cite{beal2003,bronson2009,okamoto2012,persson2013,vandemeent2013,vandemeent2014}. The
number of states, corresponding to, for example, distinct DNA-protein
conformations, is often a key quantity of interest, and the
possibility to extract it directly from the data will be especially
useful for poorly characterized and complex systems (for example, TPM
data with three rather than two operators present, as in the wild-type
{\it lac} operon \cite{johnsonPhD}).  Also in contrast with previous
methods, vbTPM handles common experimental artifacts gracefully, by
classifying them in separate states that can easily be filtered out
based on their unphysical parameters. Finally, we demonstrate further
improved resolution from an ability to pool information from large
heterogeneous data sets, using
an \mlp{EB} \sjp{approach} \cite{vandemeent2013,vandemeent2014}. Combined,
these represent significant improvements over previous analysis
methods, which we expect to be useful for a wide range of TPM
applications. Our code, implemented in a mixture of Matlab and C, is
freely available as open-source software.

\jwp{\mlp{Our} analysis of LacI-mediated loop formation
in DNA constructs with loop lengths from 100 to 109~bp \mlp{is
consistent}} with previous results \jwp{\cite{johnson2012}},
\jwp{in the sense that} we resolve
three states that cluster according to the emission parameters of the
model, $K$ and $B$, \mlp{and} which we denote the unlooped state (U),
middle looped state (M), and bottom looped state (B). \jwp{Our EB
analysis further demonstraces} that when the M and B looped states
occur in a single trajectory, they can directly interconvert without
passing through an unlooped state.  This strongly indicates that these
M and B states share a DNA binding topology but differ in LacI
conformation, because a change of DNA topology would presumably
require an unlooped intermediate, as different DNA topologies require
the unbinding and re-binding of at least one LacI DNA binding domain
from the DNA. Our finding of direct interconversions between the M and
B states are consistent with previous results on longer (138
bp \cite{wong2008} and 285 bp \cite{rutkauskas2009}) loops, which were
attributed to transitions between a V-shaped and an extended LacI
conformation.

Interestingly, at many loop lengths we can distinguish two kinds of
trajectories, those that contain both an M and a B state (which can
interconvert), and those that exhibit only one of the two looped
states (\Fig{fig:E8series}). Which of the looped states (B or M) a
two-state trajectory exhibits is the same for \sjp{essentially} all
two-state trajectories at a given loop length, but whether this state
is the M or B state varies with loop length. %The kinetics of loop
formation and
%breakdown are fast enough that such a pattern is unlikely
%to \ml{reflect insufficient equilibration of simple three-state
%kinetics}, and, 
\sj{
As discussed in the Results section and in
Sec.~\SIref{sitext:equilibration}, \sjp{for most constructs} we observe significantly more
two-state trajectories than we would expect from the null hypothesis
that this ``2+3'' pattern reflects insufficient equilibration of
simple three-state kinetics. Although we cannot conclusively rule out
the null hypothesis, we find the evidence for two different
subpopulations sufficiently compelling to propose an alternative
hypothesis, namely the existence of three different underlying loop
structures.  Taking the 2+3 pattern} together with the indication that
the single loop state changes with operator phasing
(\Fig{fig:E8series}), we \sj{argue} that this pattern reflects the
existence of two loop structures that can interconvert directly via a
conformational change in LacI, and one structure that cannot
interconvert directly to any other looped state, but has the same TPM
signature as one of the interconverting states. Interconversion
between the two- and three-state regimes is slow compared to our
typical trajectory lengths \sjp{(\Fig{fig:23pattern})}, which is the reason we can distinguish
them.

We note that a mixture of two- and three-state trajectories was also
seen in a 138-bp construct with directly interconverting looped
states, flanked by two Oid operators \cite{wong2008}.  For a 285 bp
loop flanked by two O1 operators, only trajectories with two looped
states were reported \cite{rutkauskas2009}. Closer analysis of these
data might be interesting in light of our observations.

% ML: removed the statement ``... but the two-state trajectories were
% not analyzed further'', about the Wong paper. Rereading, it seems
% they did not throw away 2-state trajectories, although they do not
% report comparing the fraction of them either.

\begin{figure*}[t!b]
\begin{center}
\includegraphics[width=\textwidth]{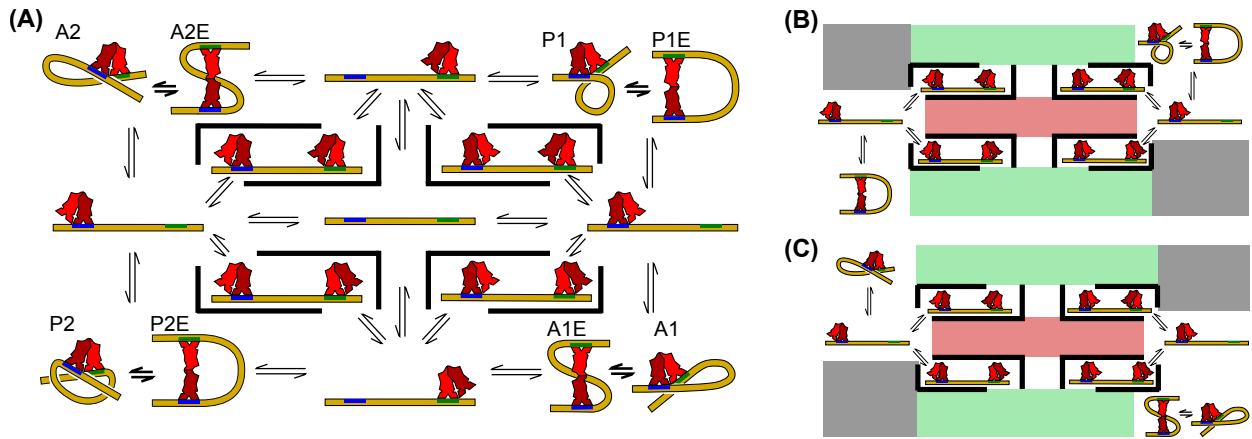}
\end{center}
  %Note that we have here drawn four different doubly occupied states,
  %which differ in the binding orientations of the two LacI's; and
  %that for clarity we have not drawn LacI in an extended conformation
  %in any of the unlooped conformations, which does not however imply
  %such extended conformations do not exist}.  
\caption{(A) 
  Loop structures arranged by LacI binding directions on the operators
  Oid (blue) and O1 (green).  \sj{These binding directions determine
  the loop topology, which, in keeping with the conventions in the
  literature, we have labeled as A1, A2, P1, and P2.}  Transitions
  between loops of different topologies (corners) are only possible
  via unlooped neighbor states\sj{, indicated by double arrows.
  However, transitions between loops that share binding topologies
  ({\it e.g.} between A1 and A1E, P1 and P1E, etc) can occur directly,
  without passing through an unlooped conformation, as we have
  demonstrated in this work, and are indicated by shorter and thicker
  double arrows}.  Singly occupied states can also interconvert
  via \sj{the} unoccupied (center) or doubly occupied states\sj{,
  which are here surrounded by thick black bars to indicate forbidden
  transitions---for example, a doubly occupied state must transition
  to a singly occupied state before a loop can form.} \ml{Note that
  extended LacI conformations may well exists also in the unlooped
  states \cite{ruben1997,taraban2008}, but for reasons of clarity, we
  have only drawn V-shaped LacI conformations in these cases, to
  highlight the different binding orientations.}  (B,C) show two
  hypothetical \sj{divisions of the} state space in (A) into two very
  slowly interconverting \sj{topology ``islands''} separated by
  energetically unfavorable states (grayed out)\sj{, low probability
  states (pink), and kinetically rare states (green); see text for
  details}. In order for these divisions to generate the observed
  2+3-state patterns of Fig.~\ref{fig:E8series}, one of the state
  ``islands'' must support only one looped state, while the other must
  support two looped states that can interconvert with one
  another. Panels (B,C) illustrate two possible ways to realize such
  behavior, in which the direct $B\ \rightleftharpoons M$
  interconversions \sj{are pictured as corresponding} to transitions
  between V-shaped and extended LacI conformations. The two different
  divisions shown in (B) and (C) might represent in-phase versus
  out-of-phase operators, which differ in which observed looped state
  (M or B) is present in \sj{two-state} trajectories (see
  Fig.~\ref{fig:E8series})\sj{.  For example, under the somewhat
  simplistic assumption here that the B state we observe by TPM always
  corresponds to an extended conformation, and the M state to a
  V-shaped conformation, then panel (B) would represent a hypothetical
  scenario for out-of-phase operators, which have two B states (one
  that interconverts with an M state, and one that does not); and (C)
  would represent in-phase operators, which have two M states (an
  interconverting one and one that does not interconvert).  Since the
  phasing of the operators determines the amount of twist in the loop,
  it is plausible that the most energetically favorable loop
  topologies would change with operator
  phasing \cite{swigon2006,towles2009,czapla2013}}. }\label{fig:lacloops}
\end{figure*}

Unraveling the structural basis for this behavior will require further
experimental, theoretical and computational efforts beyond the scope
of this paper, but it is interesting to speculate about possible
underlying molecular mechanisms. We propose as a starting point the
scheme outlined in \Fig{fig:lacloops}.
\Fig{fig:lacloops}A shows various potential loop structures
arranged by binding topology ({\it i.e.} binding direction on the
operators), with loop topology groups separated by unlooped
intermediates.  Both V-shaped and extended conformations are shown for
each group of loop topologies, \sj{and are depicted as able to
interconvert (thicker, shorter double arrows),} though it is not clear
that all \sj{topologies} are energetically feasible \sj{at the loop
lengths we study here}, nor that all \sj{loop topologies} can
convert \sj{between an extended and V-shaped conformation}.  Loop
formation and breakdown occur via transitions
\sj{to neighboring unlooped intermediates, as indicated by the thinner, 
       longer double arrows}. Singly occupied unlooped states can also
interconvert via doubly occupied intermediates. 

How could this state-space be split into two slowly interconverting
subsets as \ml{our results suggest}? First, we note that for the
operators used here, the statistical mechanics analysis from our
previous work implies that the no-LacI-bound state (center in
\Fig{fig:lacloops}A)
is essentially unpopulated at 100 pM LacI \cite{johnson2012}\sj{, so
we have eliminated it as a possible state in our system, as indicated
by the pink boxes in \Fig{fig:lacloops}B,C}. Second, we suppose,
as shown \sj{by gray boxes} in \Fig{fig:lacloops}B,C, that all
energetically feasible loops are found only in two diagonally opposite
loop topology groups, which therefore form isolated state ``islands''
separated by energetically unfavorable states.  Theoretical and
computational work consistently finds some loop topologies to be more
stable than others \cite{zhang2006,swigon2006,towles2009,czapla2013},
making this supposition tenable.  If we further hypothesize that not
all extended states can interconvert with their cognate V-shaped
topological equivalents (or vice-versa), then we would obtain the
mixture of two-state and three-state trajectories that we observe in
our data. \ml{Interconversions between two- and three-state regimes
would then be limited by the need to change LacI binding orientation
on the strong operator via multiple unlooped intermediates\sj{, which
we will argue below is sufficiently slow, given the strength of the
operators in our constructs, as to be virtually undetected on the
timescales we deal with here}}.

A final consideration for this scheme relates to the possibility of
passing from one state ``island'' to the other by way of a
doubly-occupied state.  That is, it is possible to move from
a \sj{loop topology} ``corner'' to a singly-bound neighbor state, then
to a doubly-occupied state, then to the diagonally opposite corner via
unbinding of the original LacI.  The relatively low frequency of state
transitions in our data combined with the relative dissociation rates
of LacI for the Oid and O1 operators we use here make this pathway
unlikely on the timescales of our trajectories.  Oid is about four
times stronger than
O1 \cite{johnson2012,whitson1986a,whitson1986b,hsieh1987,frank1997},
and off-rates for Oid and O1 under experimental conditions similar to
ours have been determined to be about \mlp{0.12 min$^{-1}$ and 0.3
min$^{-1}$} respectively \cite{wong2008,winter1981} (similar values
have recently been measured \textit{in vivo} as
well \cite{hammar2014}). Looped and doubly-occupied states are
therefore almost three times more likely to decay by O1 unbinding, and
so we speculate that \sj{the unlooped states covered by green boxes}
in \Fig{fig:lacloops}B and C act as \sj{kinetic} barriers between the
two outer columns. \sj{That is, we hypothesize a very slow
interconversion between the binding orientation at Oid for a given
trajectory, because unbinding from O1 is so much more
likely.} %Upperand lower loop topology groups in each column can
%easily interconvert via the singly-occupied Oid-bound state.
Moreover, recent work hints at additional types of unlooped states,
which might further slow down transitions between different topology
groups \cite{revalee2014,goodson2013}. \sj{Over long enough
timescales, though, we would imagine that a significant number of
trajectories would eventually explore both topology ``islands'' in
either \Fig{fig:lacloops}B or C, by passing through one of the green
boxes.}

The scheme we propose in \Fig{fig:lacloops} illustrates how our
results point to new interesting directions for future investigations
into LacI-mediated looping.  For example, \mlp{much} theoretical work
has focused on looping free
energies \cite{zhang2006,towles2009,czapla2013}, which are not enough
to address the question of allowed interconversions.  Another
interesting question \ml{is the possibility that great rotational
flexibility \sjp{in LacI}, of either the DNA binding domains \cite{villa2005} or the
dimers around the tetramerization
domain \cite{ruben1997,taraban2008},} might blur the differences
between loop topology groups.
%Another interesting question stems from the putative rotational
%flexibility of the DNA binding domains \cite{villa2005}, which might
%blur the differences between loop topology groups.
Finally, a computational investigation of the RMS signal for different
looped states shown in \Fig{fig:lacloops}, including the effect of the
bead and nearby \sjp{coverslip} \cite{towles2009}, would aid in matching
different structural models directly to TPM data.

%For example, it is curious that two of the three looped states we see
%have very so similar TPM signal (\Fig{fig:E8series}). Is there a
%structural reason for this?  A more thorough kinetic analysis than
%shown here would also be valuable, to extract interconversion
%kinetics and J-factors for the three-loop constructs.  To summarize,
%we have used newly developed analysis methods for TPM data to resolve
%three Lac-mediated DNA loop structures in several short constructs,
%one more than previously demonstrated.  These results provide novel
%testing grounds for mechanical theories of looping, and indicates
%that both the flexibility of the Lac tetramer and the symmetry of
%it's DNA binding domains are important for loop formation.

% I like the argument, but find the sentences too long...

Regardless of which molecular structures underlie the interconverting
and non-interconverting loop states that we observe, it is clear
that \sj{our novel Bayesian analysis was central to our ability to
resolve evidence for more than two coexisting looped states in a
single construct with TPM.} This is one more \sj{looped conformation}
than previously observed at the single molecule
level \cite{wong2008,rutkauskas2009,han2009,johnson2012,johnson2013,revalee2014},
but is in qualitative agreement with theoretical and computational
results \cite{zhang2006,swigon2006,towles2009,czapla2013} (see
Introduction).  Our findings are also consistent with recent ensemble
FRET studies with loops formed from a library of synthetic pre-bent
DNAs, in which at least three loop structures (a mixture of V-shaped
and extended) contributed significantly to the observed looping for at
least 5 of the 25 constructs examined \cite{haeusler2012}.

The impact of these different loop structures on the ability of LacI
to regulate the genes of the {\it lac} operon {\it in vivo} remains to
be seen. Theoretical work has shown that several classic features of
{\it in vivo} gene repression data with LacI can be best explained by
the presence of more than one loop conformation, and that the presence
of multiple looped states generally dampens oscillations in gene
regulation as a function of loop length \cite{Saiz2007}. Extending
these arguments, the presence of multiple looped states should allow
looping under a wider range of conditions, and hence make gene
regulation more robust against mechanical perturbations from, for
example, changes in supercoiling state or the presence versus absence
of architectural proteins.  \mlp{On the other hand}, inducer molecules
and architectural proteins such as HU have been suggested to also
change the relative stability of different loop
shapes \cite{Saiz2007,czapla2013,Becker2005,Becker2007,Becker2008,goodson2013}
which may add an additional level of regulatory potential to the
operon.

The above effects could clearly be present and relevant also
in more complex regulatory systems \sjp{of} eukaryotic cells.  A
fuller understanding of the loop structures and interconversion
pathways available to the LacI-mediated loops we observe {\it in~vitro}, and how they are influenced by architectural proteins that are
known to play a large role in gene regulation {\it in~vivo} \cite{Becker2005,Becker2007,Becker2008}, promises to greatly
enhance our understanding of this potential additional layer of gene
regulatory information.

%\section{CONCLUSION}
\paragraph{Acknowledgments}{\small 
We thank members of the Phillips lab for helpful discussions and
advice, Jason Kahn for helpful discussions about his \ml{lab's} work
in relation to ours\ml{, and the Elf and Meiners labs for sharing
  Refs.~\cite{revalee2014,hammar2014} before publication.

This work was supported by the National Science Foundation through a
graduate fellowship to S.J.; a Rubicon fellowship [grant number
  680-50-1016] from the Netherlands Organization for Scientific
Research to J.W.M.; the National Institutes of Health [grant number
  DP1 OD000217A (Director's Pioneer Award), R01 GM085286, R01
  GM085286-01S1, and 1 U54 CA143869 (Northwestern PSOC Center)], and
the Foundation Pierre Gilles de Gennes to R.P.; an NIH National
Centers for Biomedical Computing grant (U54CA121852) to C.H.W.; the
Wenner-Gren foundations, the foundations of the Royal Swedish Academy
of Sciences, and the Foundation for strategic research (SSF) via the
Center for Biomembrane research to M.L.}

\paragraph{Conflict of interest statement.} None declared.\\
%\noindent Supplementary Data\ml{, available at NAR online, include
%  supporting text and raw TPM traces for the E8106 and E8107
%  constructs}.

%For updated versions of the vbTPM suite, see vbtpm.sourceforge.net.

%\bibliographystyle{narML} \bibliography{references}

\clearpage
\section*{Supporting information}\addcontentsline{toc}{section}{Supporting information}
% introduce S before equations, figures, and tables in the SI.
\renewcommand{\theequation}{S\arabic{equation}}
\renewcommand{\thesection}{S\arabic{section}}
\renewcommand{\thesubsection}{S\arabic{section}.\arabic{subsection}}
\renewcommand{\thefigure}{S\arabic{figure}}
\renewcommand{\thetable}{S\arabic{table}}
\setcounter{section}{0}
\setcounter{equation}{0}
\setcounter{figure}{0}
\tableofcontents
\addtocontents{toc}{\protect\setcounter{tocdepth}{2}}
%!TEX root = HMM_SI.tex
\section{vbTPM work{f}{l}ow}\label{sisec:workflow}

\begin{figure}[!t]
\begin{center}
  \includegraphics[width=2.5in]{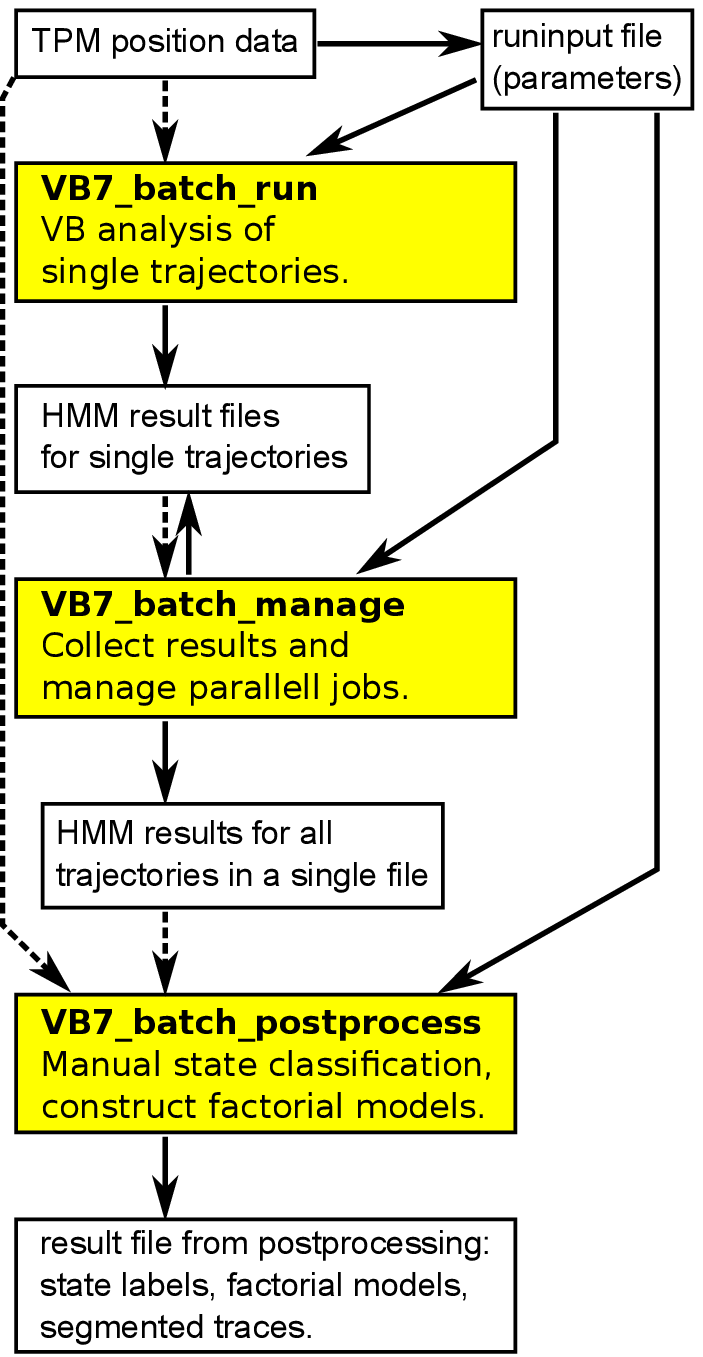}
  \end{center}
  \caption{Work flow for TPM analysis using vbTPM. The yellow boxes
    indicate the three main tools of the vbTPM toolbox and their
    functions, as described in the text. Solid lines indicate that a
    file is written by another file, or passed as argument to
    it. Dashed lines indicate flow of information handled internally
    by reference to the runinput file.}\label{sifig:analysisflow}
\end{figure}

The workflow of vbTPM, summarized in \Fig{sifig:analysisflow}, is
based on runinput files that contain all analysis parameters,
including information about where the TPM data files are located, and
where various results should be written to. These files can therefore
be used as handles to an ongoing analysis and to intermediate results.

The three main tools for handling the analysis, marked in yellow in
\Fig{sifig:analysisflow}, are \\
\noindent{\bf VB7\_batch\_run.m}, which manages the VB analysis of raw
position traces using the simple HMM model,\\
\noindent{\bf VB7\_batch\_manage.m}, a tool to collect the analysis
results, and also to clean up and reset intermediate result files in
case the analysis is interrupted, and finally\\
\noindent{\bf VB7\_batch\_postprocess.m}, a graphical tool to aid the
manual state classification and construct factorial models based on
this classification.

More advanced analysis beyond this step, including the EB procedure,
currently require custom Matlab scripting.  Further details are given
in the software manual.

\section{The emission parameters}\label{sec:RMSderivation}
To gain more physical intuition about the parameters $K,B$ that model
the bead motion, we derive the corresponding expressions for the
standard deviation (or RMS value) and the correlation time, for the
case with no hidden states. The bead motion model, \Eq{M:eq:eomx} in
the main text, can be expressed as a stochastic difference equation
whose parameters depend on the hidden state,
\begin{equation}
  \x_t=K_{s_t}\x_{t-1}+\vec{w}_t/(2B_{s_t})^{1/2},
\end{equation}
where $\vec{w}_t$ are independent vectors of Gaussians with two
independent components and unit variance, 
\begin{equation}
  \mean{w_t^{(i)}w_{u}^{(j)}}=\delta_{t,u}\delta_{i,j},\quad i,j=x,y.
\end{equation}
With no hidden states, this simplifies to
\begin{equation}\label{seq:eomx1}
  \x_t=K\x_{t-1}+\vec{w}_t/(2B)^{1/2},
\end{equation}
and to compute the corresponding RMS value, we first substitute
\Eq{seq:eomx1} into $\mean{\x_t^2}$, to get
\begin{multline}
  \mean{\x_t^2}
  =\mean{\left(K\x_{t-1}+\frac{\vec{w}_t}{\sqrt{2B}}\right)^2}\\
  =K^2\mean{\x_{t-1}^2}
  +\frac{ \mean{\vec{w}_t^2}}{2B}
  +\sqrt{\frac2B}\mean{\vec{w}_t\cdot\x_{t-1}}.
\end{multline}
Now, the equation of motion \eqref{seq:eomx1} means that $\vec{w}_t$
and $\x_{t-1}$ are independent, so that
$\mean{\vec{w}_t\cdot\x_{t-1}}=0$, and since $\x_t$ is also stationary
it follows that $\mean{\x_t^2}=\mean{\x_{t-1}^2}=RMS^2$. Finally, noting
that $\mean{\vec{w}_t^2}=\mean{(w_t^{(x)})^2}+\mean{(w_t^{(y)})^2}=2$,
the equation for $\mean{\x_t^2}$ simplifies to 
\begin{equation}
  \mean{\x_t^2}=K^2\mean{\x_t^2}+1/B,
\end{equation}
which leads to the expression for the RMS value of \Eq{M:eq:1state},
\begin{equation}
  RMS=\sqrt{\mean{\x_t^2}}=\big(B(1-K^2)\big)^{-1/2}.
\end{equation}

To derive the correlation time, we similarly start with the equation
of motion \eqref{seq:eomx1} to compute
$\mean{\x_t\cdot\x_{t-1}}$. After applying the same type of arguments,
we get
\begin{multline}
  \mean{\x_t\cdot\x_{t-1}}
  =\mean{\left(K\x_{t-1}+\frac{\vec{w}_t}{\sqrt{2B}}\right)\cdot\x_{t-1}}\\
  =K\mean{\x_{t-1}^2}+0=K\mean{\x_t^2}.
\end{multline}
Repeated application to longer times, and division by $\mean{\x_t^2}$,
leads to
\begin{equation}
  \frac{\mean{\x_{t}\cdot\x_{t-m}}}{\mean{\x_t^2}}=
  \frac{\mean{\x_{t+m}\cdot\x_{t}}}{\mean{\x_t^2}}=K^{|m|}\equiv
  e^{-|m|\Delta t/\tau},
\end{equation}
where the last step is just the definition of the correlation time
$\tau$ in terms of the sampling time $\Delta t$. This is indeed the
correlation time given in \Eq{M:eq:1state}.

\section{Choice of priors}
We would like to choose uninformative prior distributions in order to
minimize statistical bias.  This is unproblematic for the emission
parameters $K,B$, since the amount of data in all states is large
enough to overwhelm any prior influence. As derived in the software
manual\footnote{See vbtpm.sourceforge.net for the latest version.},
prior distributions for $K,B$ are given by
\begin{align}\label{seq:KBtrial}
  p(\vec K,\vec B|N)=&\prod_{j=1}^N
  \frac{B_j^{\tilde n_j}}{W_j}e^{-B_j\big(\tilde v_j(K_j-\tilde \mu_j)^2+\tilde c_j\big)},\\
  W_j=&\frac{\tilde c^{-(\tilde n_j+\frac 12)}
    \Gamma(\tilde n_j+\frac 12)}{\sqrt{\tilde v_j/\pi}},
\end{align}
with the range $B_j\ge 0$, $-\infty<K_j<\infty$.  Throughout this
work, we use
\begin{align}
  \tilde\mu_j=&0.6, &
  \tilde n_j=&1,\\
  \tilde v_j =&5.56\text{ nm$^2$},&
  \tilde c_j=&30000\text{ nm$^2$},
\end{align}
which corresponds to 
\begin{align}
  \mean{K_j}=&0.6,&
  \mean{B_j}=&5\times 10^{-5}\text{ nm$^{-2}$},\\
  \mathrm{std}(K_j)=&0.3,&
  \mathrm{std}(B_j)=&141.4\times 10^{-5}\text{ nm$^{-2}$}.
\end{align}

The prior for the initial state probabilities are Dirichlet
distributed, $p(\vec\pi|N)=\Dir(\vec\pi|\tilde{\vec{w}}^{(\vec\pi)})$,
and these variables are unproblematic for the opposite reason: the long
length of the trajectories makes the initial state relatively
unimportant to describe the data. We use a constant prior of strength
5, i.e.,
\begin{equation}
  \tilde w_j^{(\vec\pi)}=5/N,
\end{equation}
where $N$ is the number of hidden states.

The transition probabilities need more care, because the potentially
low number of transitions per trajectory makes the prior relatively
more influential.  The prior for the transition matrix $\matris{A}$
are independent Dirichlet distributions for each row, parameterized by
a pseudo-count matrix $\tilde w_{ij}^{(\matris{A})}$.  Following
Ref.~\cite{persson2013}, we parameterize this prior in terms of an
expected mean lifetime and an overall number of pseudo-counts (prior
strength) for each hidden state.  In particular, we define a
transition \textit{rate} matrix Q with mean lifetime $t_D$,
\begin{equation}
  Q_{ij}=\frac{1}{t_D}\left(-\delta_{ij}
  +\frac{1-\delta_{ij}}{N-1}\right),
\end{equation}
and then construct the prior based on the transition probability
propagator per unit timestep,
\begin{equation}
  \tilde w_{ij}^{(\matris{A})}=\frac{t_Af_{sample}}{n_{downsample}}
  e^{\Delta tQ}.
\end{equation}
Here, $t_A$ is the prior strength; both $t_A$ and $t_D$ are specified
in time units to be invariant under a change of sampling
frequency. Further, the timestep is given by $\Delta
t=n_{downsample}/f_{sample}$, where $f_{sample}$ is the sampling
frequency (30 Hz in our case), and $n_{downsample}$ is the
downsampling factor (we use 3).

Numerical experiments in Ref.~\cite{persson2013} show that choosing the
strength too low compared to the mean lifetime produces a bias towards
sparse transition matrices. This is not desirable in our case, and we
therefore use $t_D=1$ s, and $t_A=5$ s throughout this work.

%!TEX root = HMM_SI.tex
\section{Performance on synthetic data}

\begin{figure}[!tb]
\includegraphics[width=\columnwidth]{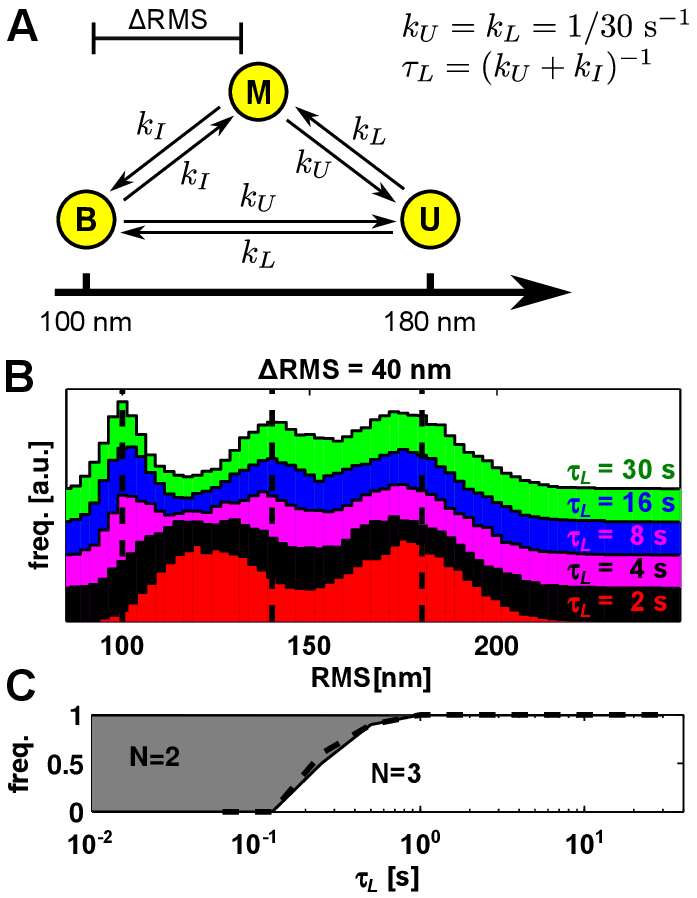}
\caption{Temporal resolution with vbTPM and RMS histograms.  (A) Model
  for synthetic data, with the difficulty determined by the
  RMS-separation $\Delta$RMS and mean life time $\tau_L$ of the two
  interconverting states M and B.  (B) Aggregated RMS histograms from
  ten 45-min trajectories with $\Delta$RMS=40 nm and varying
  $\tau_L$. The M and B states are blurred to a single peak at low
  $\tau_L$, but for $\tau_L\gtrsim 8$ s, all three states can be
  resolved.  Vertical lines show the true RMS values. (C) Fraction of
  trajectories in which the HMM algorithm resolved 2 (gray) or 3
  (white) states. All three states are resolvable at $\tau_L\ge
  0.5$ s, significantly better than the histogram method. The dashed
  line shows the result without downsampling, an insignificant
  improvement.  The filter width used in (B) was optimized by eye to
  $\sigma_G=3$ s.}\label{sifig:resolution}
\end{figure}

Here, we test the abilities of vbTPM to resolve close-lying states in
synthetic data, and compare it to the RMS histogram method. We also
verify that model parameters are recovered correctly, and that these
results are insensitive to the factor three downsampling that we use
for analysis on real data.

Our test model, depicted in Fig.~\ref{sifig:resolution}A, has one
unlooped (U) and two looped (M and B) states, and the difficulty of
resolving states M and B can be tuned by decreasing either their RMS
difference $\Delta$RMS or their average life-time $\tau_L$ (the
life-time of the aggregated state B+M is fixed at 30 s, same as the
unlooped state). For each parameter setting, we generated and analyzed
ten 45 minute trajectories.

\Fig{sifig:resolution}B-C shows a comparison of temporal resolution,
using $\Delta$RMS=40 nm and varying $\tau_L$. Resolving states using
histograms means resolving peaks, and three distinct peaks emerge at
$\tau_L=4-8$ s. In contrast, vbTPM resolves the correct number of
states already at $\tau_L=0.5$ s.  This order-of-magnitude improvement
mainly reflects the detrimental effects of the low-pass filter used in
the RMS analysis, and is insensitive to downsampling by a factor of 
three. The vbTPM limit can instead be compared to the bead correlation
time $\tau$, which were set to 0.1, 0.17, and 0.25 s for the B,M and U
states in this data.

We also compared vbTPM to the histogram method for resolving states
that interconvert slowly ($\tau_L=30$ s) with varying degrees of
separation in RMS. The result is shown in \Fig{sifig:RMSresolve}, and
indicates that vbTPM does not significantly outperform the histogram
method in this case.

To summarize, we mapped out the resolution of vbTPM in the range
0.0625 s $\le \tau_L \le $ 30 s, 5 nm $\le \Delta RMS \le$ 40 nm. The
results, in \Fig{sifig:HMMresolution}, show a nonlinear relation
between the spatial and temporal resolution.

Next, we verify that model parameters are also well reproduced and
insensitive to downsampling in this situation.  \Fig{sifig:dRMS40rms}
shows the RMS values for the most likely models fitted to the test
data set of \Fig{sifig:resolution}.  The looped state of the two-state
models display an average of the two looped states in the data when
those states interconvert too quickly to be resolved. The three-state
models generally reproduce the input parameters with a slight downward
bias that is more noticeable at high RMS values. We believe that this
is an effect of the drift-correction filter we applied to the
data. Note that the results with and without downsampling are almost
indistinguishable.

The mean lifetimes (\Fig{sifig:dRMS40dwell}) show similar trends of
good fit and almost no difference with and without
downsampling. Two-state models that do not resolve the two looped
states learn their aggregated mean lifetime, which is indeed 30 s in
the true model. The tendency to overestimate the short lifetimes can
be rationalized by noting that short sojourns are more difficult to
resolve, and therefore do not contribute as much to the estimated
parameter values.
 
Individual transition probabilities (elements $A_{ij}$) are presented
in \Fig{sifig:dRMS40Aij}. Here there is a clear difference with and
without downsampling, since the latter estimates transition
probability per timestep, while the former per three timesteps. Low
transition probabilities suffer significant fluctuations due to small
number statistics, while the higher transition probabilities are well
reproduced.
\begin{figure}[!h]
  \includegraphics[width=\columnwidth]{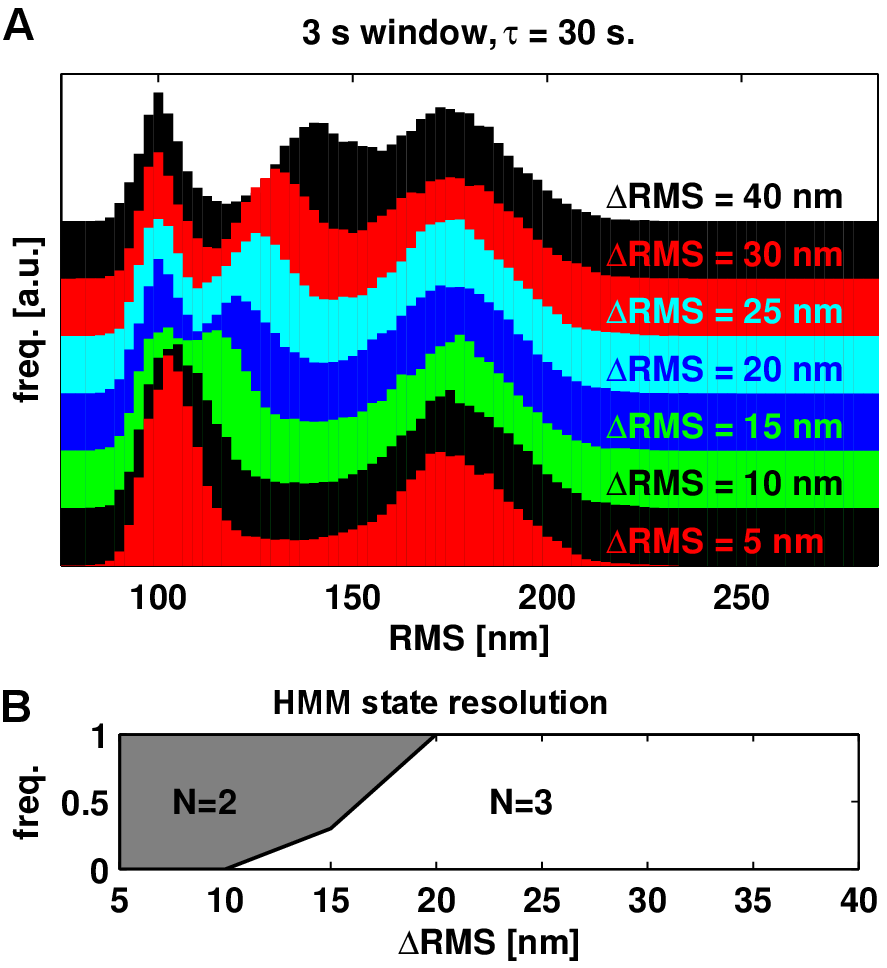}
  \caption{Resolving three states with varying $\Delta$RMS and looped
    mean life-time $\tau=30$ s. (A) Aggregated histograms for ten 45
    min-trajectories, filtered with $\sigma_G=3$ s. (B) Fraction of
    detected two- (gray) and three-state (white) models with vbTPM
    applied to the same ten trajectories one by
    one.}\label{sifig:RMSresolve}
\end{figure}

\begin{figure}[!tb]
  \includegraphics[width=\columnwidth]{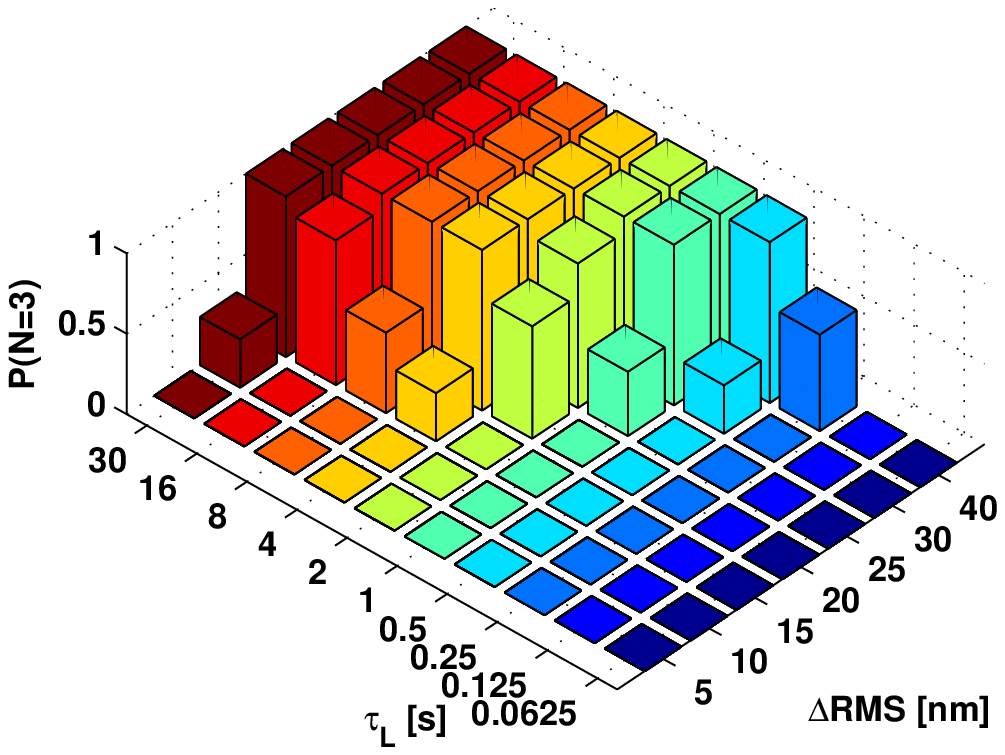}
  \caption{Resolution map of vbTPM shown as the fraction of correctly
    identified 3-state models at different $(\Delta
    RMS,\tau_L)$-pairs. Ten 45 min-trajectories were simulated at each
    parameter set, and 3-fold downsampling was used for the
    analysis.}\label{sifig:HMMresolution}
\end{figure}

\begin{figure*}[tbp]
  \includegraphics[width=\textwidth]{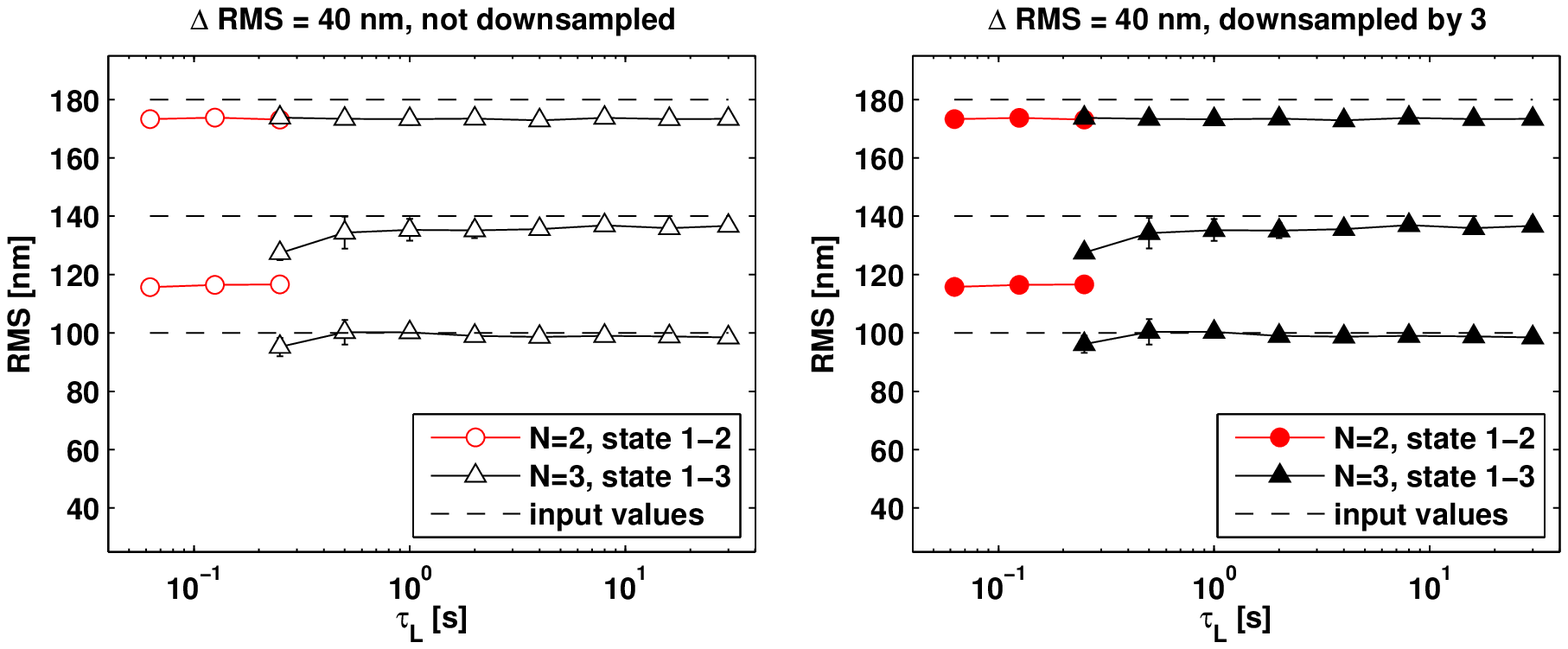}
  \caption{RMS values for the best fit models (symbols) to the data
    set in \Fig{sifig:resolution}, compared to simulated parameters
    (dashed). Posterior mean value $\pm$ std.~ (an estimate of the
    parameter uncertainty) for two- and three-state models shown
    separately, according to which model size got the best score for
    each trajectory. Most error bars are smaller than the symbols.
    Analysis without (left) and with (right) downsampling give almost
    identical results.}\label{sifig:dRMS40rms}
\end{figure*}

\begin{figure*}[tbp]
  \includegraphics[width=\textwidth]{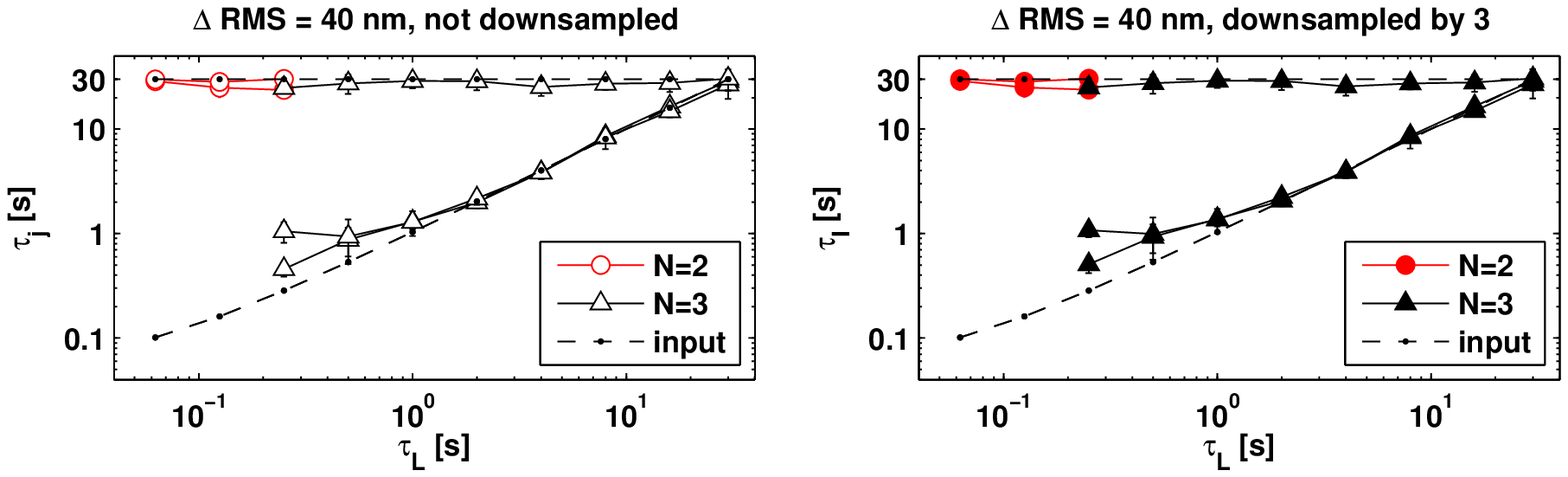}
  \caption{Mean lifetimes, presented as in \Fig{sifig:dRMS40rms}. The
    true model (dashed) has one state with mean lifetime 30 s (U), and
    two states (M and B) with shorter lifetimes. The lower dashed line
    is not straight because $\tau_L$ is defined as a rate in a
    continuous time model, while lifetimes (true and fitted) are
    defined in a discrete-time setting using the transition
    probability matrix $A_{ij}$, which makes a difference for short
    lifetimes. The average lifetime of the short-lived states together
    is always 30 s however, which explains why the two-state models
    that do not resolve these two states have both lifetimes around 30
    s.}\label{sifig:dRMS40dwell}
\end{figure*}

\begin{figure*}[tbp]
  \includegraphics[width=\textwidth]{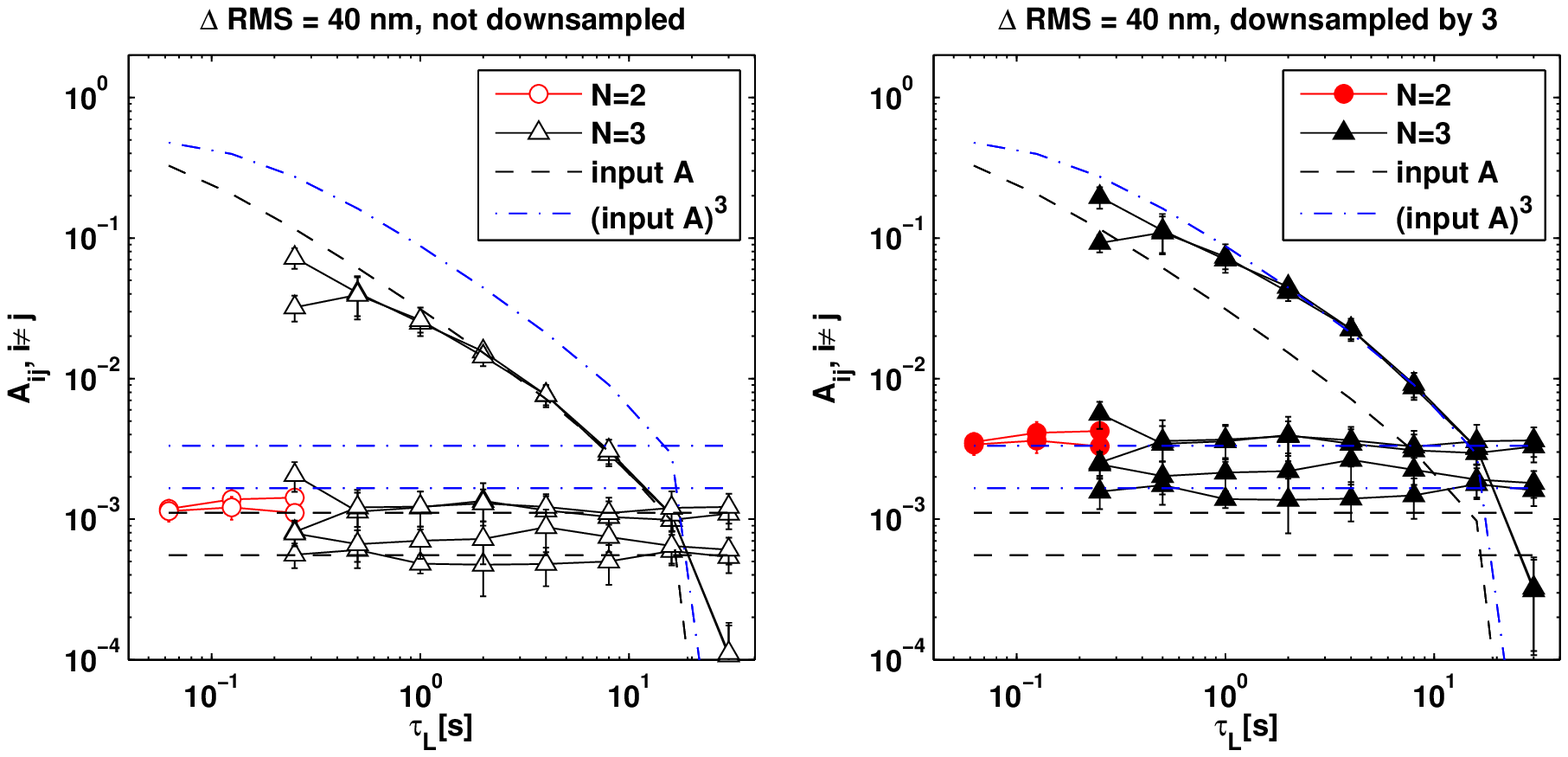}
  \caption{Transition probabilities (non-diagonal elements of
    $A_{ij}$), presented as in \Fig{sifig:dRMS40rms}. Due to
    symmetries of the underlying kinetic model it only contains three
    distinct transition probabilities. The difference with and without
    downsampling is due to the fact that the downsampled model
    effectively estimates transition probabilities per three
    timesteps, given by $\matris{A}^3$ (blue dash-dotted lines),
    instead of the single-step probabilities (black dashed lines) used
    to produce the data. Relative to these different targets, however,
    the analyses with and without downsampling give very similar
    results. }\label{sifig:dRMS40Aij}
\end{figure*}

%\FloatBarrier
\section{Effect of short-lived spurious states}\label{sec:spurious}
vbTPM is able to detect many short-lived spurious states that cannot
be detected in RMS trajectories, and one might wonder if the presence
of these states poses a problem for earlier results where they were
not detected \cite{johnson2012}. To test this, we compute some
properties of our E8 constructs subjected to our standard screening
process \cite{johnson2012} (which does not detect short-lived
artifacts), and compare them to a population where the trajectories
are subjected to additional screening, namely, where trajectories with
the most frequent short-lived spurious events are removed. The
differences turn out to be small.

For this additional screening, we looked at the average frequency of
transitions from genuine to spurious states and the fraction of time
spent in spurious states. As shown in Fig.~\ref{sifig:spuriousstats},
the distributions of these properties for the E8x and TAx trajectories
have distributions that are fairly broad.  For this comparison, we set
thresholds of at most 6 spurious transitions per minute and 5\%
spurious occupancy (dashed lines in
Fig.~\ref{sifig:spuriousstats}A,B), which removed about 30\% of all
trajectories (although the fraction varied significantly between
different constructs).

Figure \ref{sifig:spuriousdiff} shows average state occupancies, mean
dwell times, and average rates of loop-loop interconversions computed
from models converged with the EB algorithm on all trajectories in our
E8x constructs (assuming simple three-state kinetics, and should thus
be interpreted with care). Solid lines show results for trajectories
passing the standard screening, while dashed lines represent the
results after the additional screening to remove trajectories with
many short spurious events.  As seen in Fig. \ref{sifig:spuriousdiff},
the presence or absence of these ``most spurious'' trajectories
generally have a small effect on the analyzed average properties.

\begin{figure}%[!bhp]
  \includegraphics[width=\columnwidth]{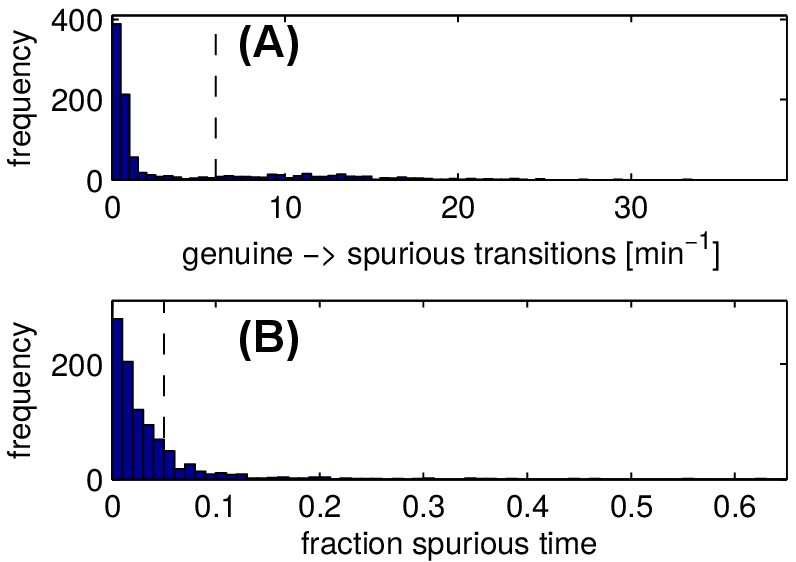}
  \caption{Distribution of short-lived spurious states in trajectories
    from all (E8x and TAx) constructs. (A) Average frequency of
    transitions from a genuine to a spurious state. (B) Fraction of
    time spent in a spurious state.}\label{sifig:spuriousstats}
  \end{figure}

\begin{figure}[!b]
  \includegraphics[width=\columnwidth]{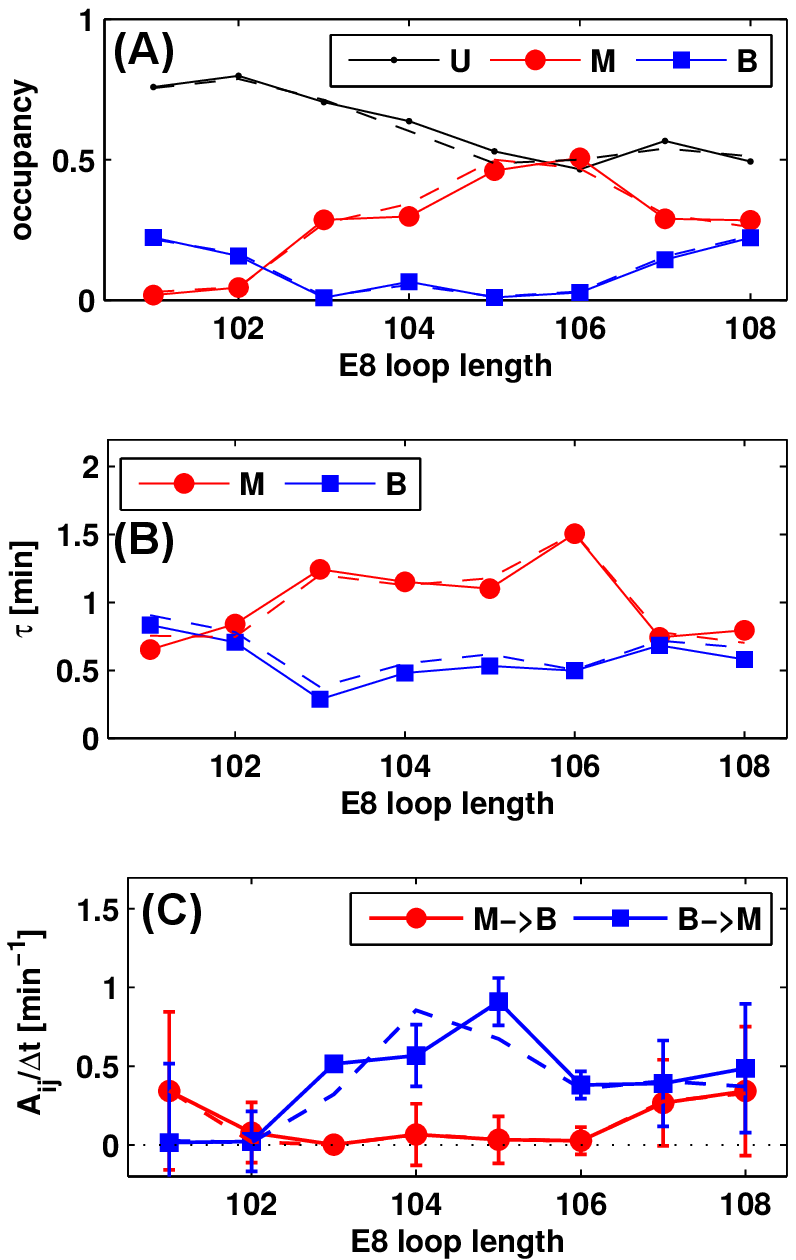}
  \caption{Comparison of (A) state occupancy, (B) mean dwell times,
    and (C) average transition probabilities between looped states for
    all E8 trajectories that passed our standard screening (solid) and
    the additional thresholds defined in this section (dashed).  Error
    bars in (C) are standard deviations. Note that the occupancy
    values in (A) are not directly comparable to those in our earlier
    analysis \cite{johnson2012}, since the effect of trajectories
    without looping activity was not corrected for in this
    plot.}\label{sifig:spuriousdiff}
\end{figure}

%!TEX root = HMM_SI.tex

\section{Equilibration analysis}\label{sitext:equilibration}

\begin{figure}[!tb]
	\includegraphics{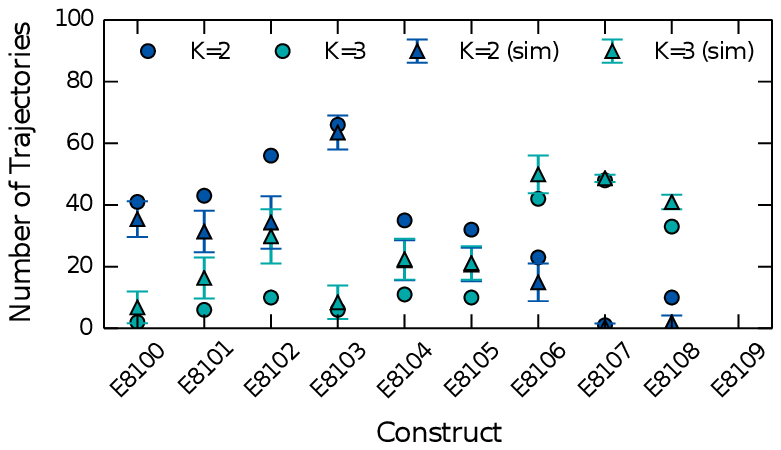}
	\includegraphics{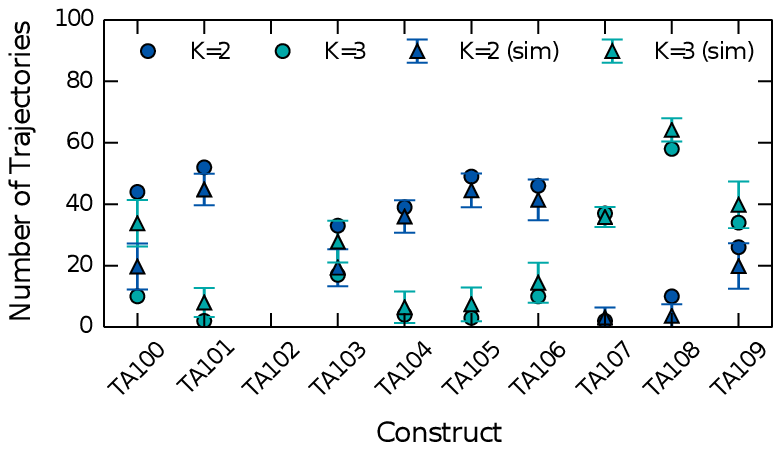}
	\caption{
	\label{fig:equilibration}
	Comparison of the number of 2-state (blue) and 3-state (cyan)
        trajectories obtained in vbTPM analysis of experimental data
        (circles) to those in simulated datasets (triangles), for E8x
        (top) and TAx (bottom) constructs. Error bars mark two
        standard deviations over 100 simulated datasets. Data for the
        TA102 and E8109 constructs are missing, because the VB results
        for those constructs showed too large variability for manual
        classification.}
\end{figure}	

As discussed in the main text, analysis of both E8x and TAx
experiments shows that a significant number of trajectories populate
only one of two looped states, along with the unlooped state, whereas
others populate all three states.  One possible explanation for the
apparent existence of 2-state and 3-state populations is that we are
simply observing equilibration effects, and that every trajectory
would eventually populate all three states, provided a bead is
observed over a sufficiently long measurement interval.  In order to
test this null hypothesis---that is, the hypothesis that all
trajectories observed for a particular construct are actually drawn
from a single, three-state population, and some of them end up only
exploring one of the two looped states due to the finite observation
time---we have generated datasets consisting of simulated state
trajectories, drawn from an underlying 3-state population, and
compared the number of 2-state and 3-state trajectories in the
simulated data to those found in the analysis of the experimental
data.

The procedure for this analysis is as follows. We first perform vbTPM
analysis of the experimental data, and then use the EB analysis to
estimate a distribution $p(A \,|\, \alpha)$ over the transition rates
and a distribution $p(\pi \,|\, \rho)$ over the initial state
probabilities. (As shown in Fig.~\ref{fig:ExTrj1}-\ref{fig:ExTrj4}
below, the EB analysis tends to give more accurate state assignments
than the VB analysis, since it uses information from multiple
trajectories at once. It also describes the variability between
individual beads.)  Note that this EB analysis implicitly assumes all
trajectories belong to a single 3-state population, though not all
trajectories are required to populate each of the 3 states.  For each
trajectory $n = 1 \ldots N$ in the experiment, we now simulate a
trajectory $s_{n,t}$ with a number of time points $T_n$ that is
identical to that of the $n$-th trajectory in the experimental data.
To do so, we first sample $A_n \sim p(\cdot \,|\, \alpha)$ and $\pi_n
\sim p(\cdot \,|\, \rho)$.  We then sample $s_{n,1} \sim p(\cdot \,|\,
\pi_n)$ and $s_{n,t} \sim p(\cdot \,|\, A_{s_{n,t-1}})$ for $t = 2
\ldots T_n$.  We repeat this procedure 100 times, using new values
$A_n$ and $\pi_n$ on each sweep.

Figure~\ref{fig:equilibration} shows the number of 2-state and 3-state
trajectories obtained through an EB analysis of real data as compared
to the corresponding numbers in simulated datasets.  In this analysis
we define a trajectory as having 3 states when $\sum_t E[s_{n,t,k}] >
5$ for all 3 states $k$.  In other words, a trajectory must assign at
least 5 time points to each state in order to be classified as having
3 states.  This threshold was empirically chosen to exclude instances
where a brief transition to a spurious state may be misinterpreted as
a transition to an actual state.  However, we verified that analysis
results were not qualitatively different when this threshold was
lowered to 1 time point for each state. Note that the EB analysis can
sometimes find a third genuine state that the VB algorithm missed, as
shown in \Fig{fig:ExTrj5}, and thus TA105 does show a few three-state
trajectories.

Analysis of the E8x trajectories shows a significantly lower number of
3-state trajectories than in equivalent simulated data.  The TAx
constructs show a similar, if less pronounced, trend; we believe that
this is due to the poor statistics for these constructs, in which the
2+3 pattern is less extreme (that is, fewer TAx constructs have a
robust mixture of 2- and 3-state populations, compared to the E8x
constructs).  Note that for the TAx constructs that do have a
significant number of both 2- and 3-state trajectories ({\it e.g.},
TA100, TA103, TA106, TA109), the trend follows that of the E8x
constructs, with more 2-state trajectories observed experimentally
than in simulated data.  Note also that the error bars on simulated
counts show an interval of two standard deviations (95\% confidence),
which represents a very conservative estimate of the uncertainty.  In
other words, under the null hypothesis where all trajectories are
described by a single 3-state model, we would expect to see a
significantly higher number of 3-state trajectories than we actually
observe experimentally, suggesting that the 2-state trajectories and
the 3-state trajectories in our data are not drawn from the same
underlying population.  These results lead us to hypothesize that we
are observing three looped states, not two, as detailed in the main
text.

\section{Detecting less rare interconversions}

\begin{figure}[t!]%[tbp]  
  \begin{center}
    \includegraphics{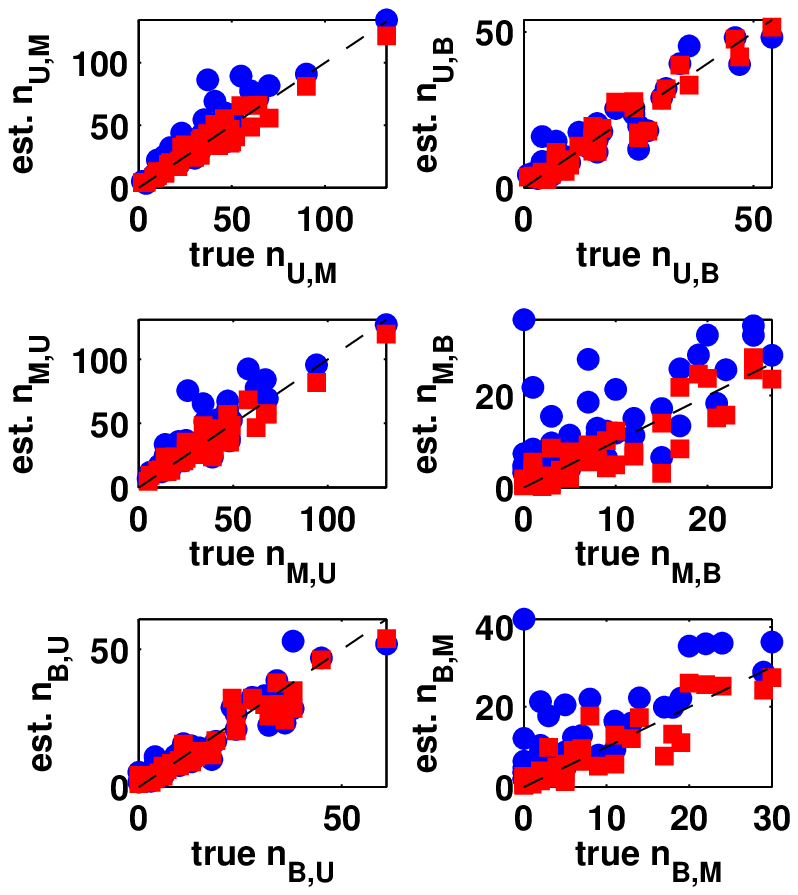}
  \end{center}
  \caption{Counting the number of interconversion in synthetic data
    based on the E8107 parameters with all rates doubled to increase
    the number of events. VB and EB results are shown in blue and red
    respectively.}\label{sifig:nMBE8107doublerate}
\end{figure}

\Fig{M:fig:nMB}A shows that the EB algorithm clearly undercounts the
number of $B \leftrightharpoons M$ interconversions in the synthetic
data based on the E8106 trajectories, where such transitions are very
rare. This downward bias is significantly smaller for the synthetic
based on E8107 parameters where these transitions are less rare. To
see if this trend continues with increasing number of events, we
generated and analyzed synthetic data based on E8107 parameters, but
with transition rates doubled. This increases the number of events
without making dwell times too short.  \Fig{sifig:nMBE8107doublerate}
shows the true and estimated counts for all types of transitions in
this data set.  Both methods work well on $U \leftrightharpoons B$
interconversions, which have the largest RMS difference, but the VB
method shows a clear bias on the less well-separated $U
\leftrightharpoons M$ and $B \leftrightharpoons M$ interconversions,
with greater bias in the latter case, where there are fewer
transitions. The EB method appears unbiased in all cases, indicating
that the tendency of EB to undercount transitions in the synthetic
E8106 data is indeed an effect of rare transitions rather than a
systematic downward bias.

\section{Loop-loop interconversions in all constructs}
Having established in \Fig{M:fig:nMB} that the EB algorithm can
reliably detect direct loop-loop interconversions, we present the
corresponding analysis for our other constructs.  For every construct,
we ran EB analysis on all trajectories identified as three-state by
the VB algorithm and manual classification, and counted the posterior
expectation of the number of BM-interconversions. However, TA105 had
no three-state trajectories in this analysis (\Fig{M:fig:E8series}),
so for this construct we instead counted BM-transitions in three-state
trajectories based on the EB analysis done for \Fig{fig:equilibration},
  an analysis that includes all trajectories. The results are shown in
  Figs.~\ref{sifig:E8interconversions}-\ref{sifig:TAinterconversions}.

The EB analysis detects loop-loop interconversions in all
constructs. However, since the total number of three-state
trajectories tend to decrease with decreasing loop length, the
evidence is most convincing for the longer constructs.

\begin{figure*}[h]
  \begin{center}
  \includegraphics[width=\textwidth]{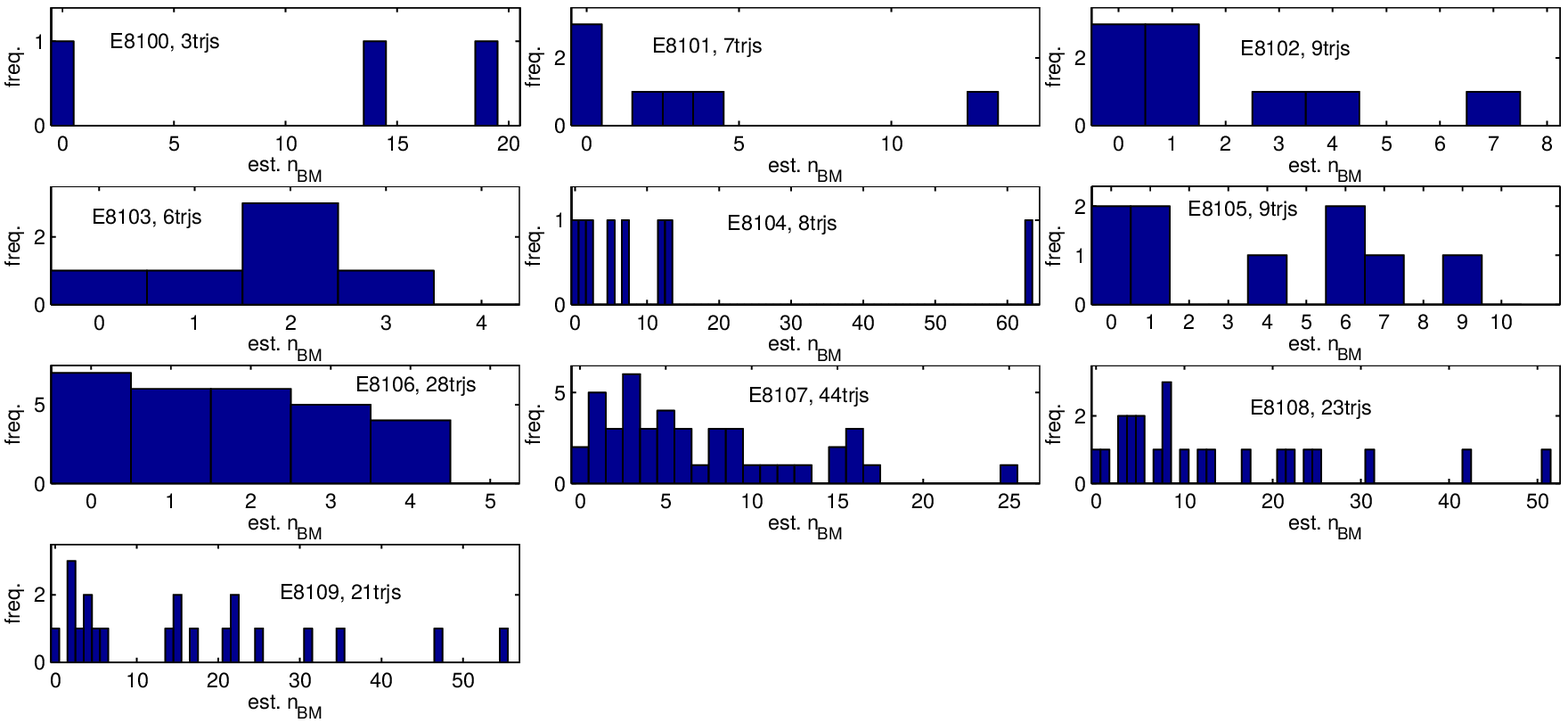}
  \end{center}
\caption{Number of loop-loop interconversions from an EB analysis of
  three-state trajectories of the E8 constructs, analogous to
  \Fig{M:fig:nMB}E,F.}\label{sifig:E8interconversions}
\end{figure*}

\begin{figure*}[b]
  \begin{center}
  \includegraphics[width=\textwidth]{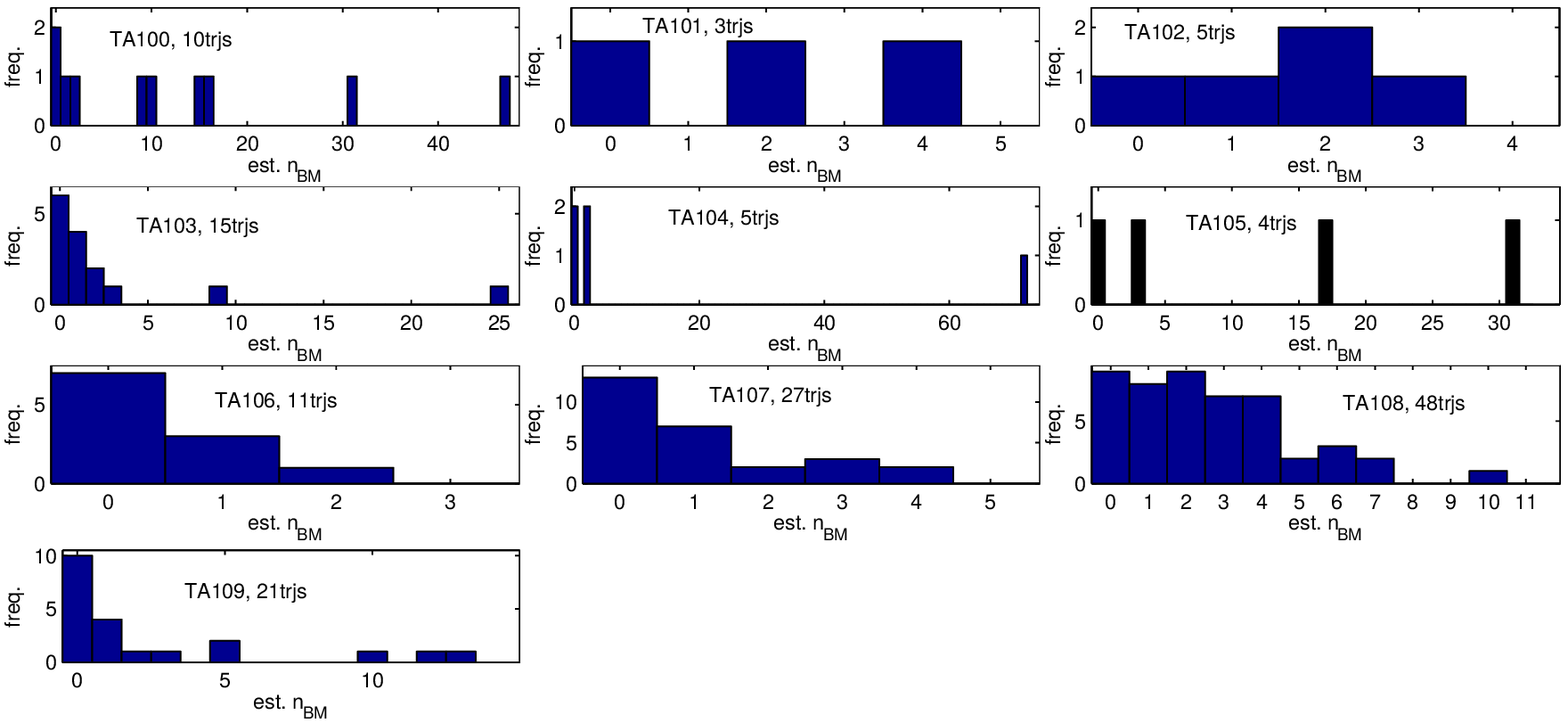}
  \end{center}
\caption{Number of loop-loop interconversions from an EB analysis of
  three-state trajectories of the TA constructs analogous to
  \Fig{M:fig:nMB}E,F. For TA105, which lacks 3-state trajectories in
  the VB analysis, we use three-state trajectories from the EB
  analysis of \Fig{fig:equilibration}.
}\label{sifig:TAinterconversions}
\end{figure*}

\section{Example trajectories}
In Figs.~\ref{fig:ExTrj1}-\ref{fig:ExTrj4}, we provide a few examples
of analyzed trajectories from the E8106 construct. Each example
shows the RMS trace (black), the sequence of most likely hidden states
from the VB analysis (``HMM'', yellow),  the sequence of most
likely genuine states from the corresponding factorial model
(magenta), and the sequence of most likely states from the
  empirical Bayes (EB) algorithm, converged with three genuine states
  on all two- and three-state trajectories (cyan). In some cases, we
also show short sections of drift-corrected position traces
($x(t),y(t)$ in blue and red), where the segmentation indicated the
presence of short-lived spurious states.

%
%\begin{figure*}
%  \includegraphics{SI_figures/E8106_1_1.eps}
%  \caption{}\end{figure*}
%
%\begin{figure*}
%  \includegraphics{SI_figures/E8106_14_1_segmentation.eps}
%  \caption{}\end{figure*}

\begin{figure*}
  \includegraphics{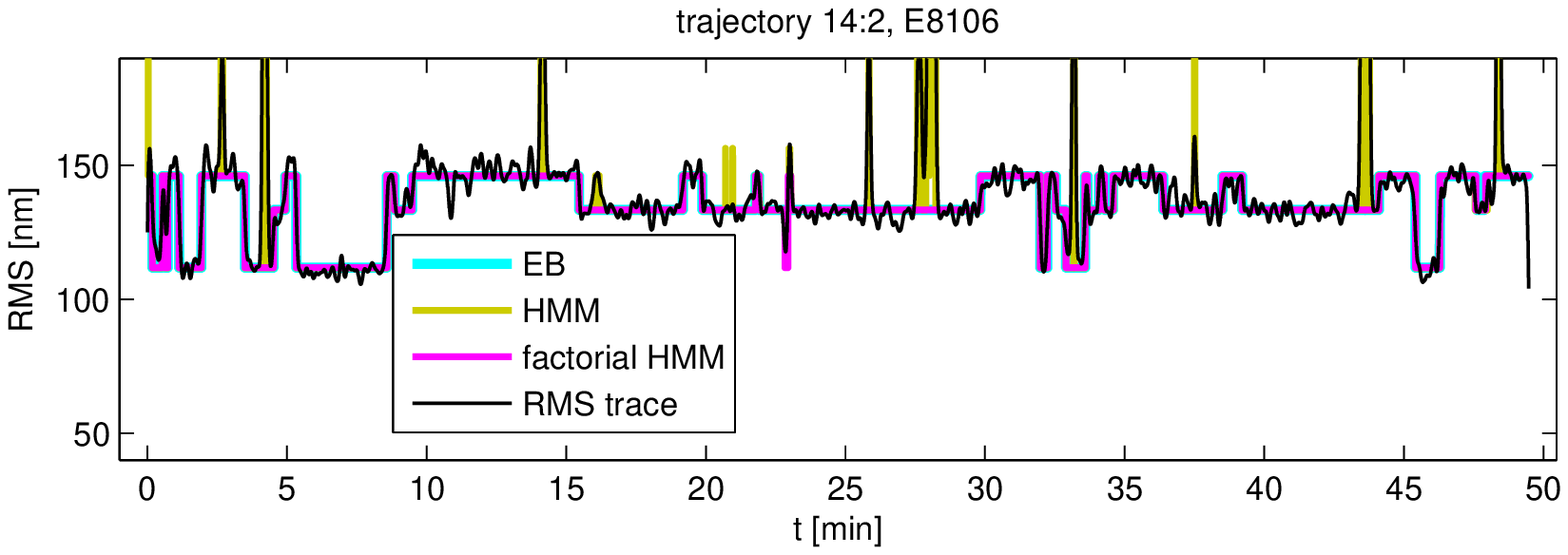}
  \caption{An example of a long, three-state trajectory.}\label{fig:ExTrj1}\end{figure*}

\begin{figure*}
  \includegraphics{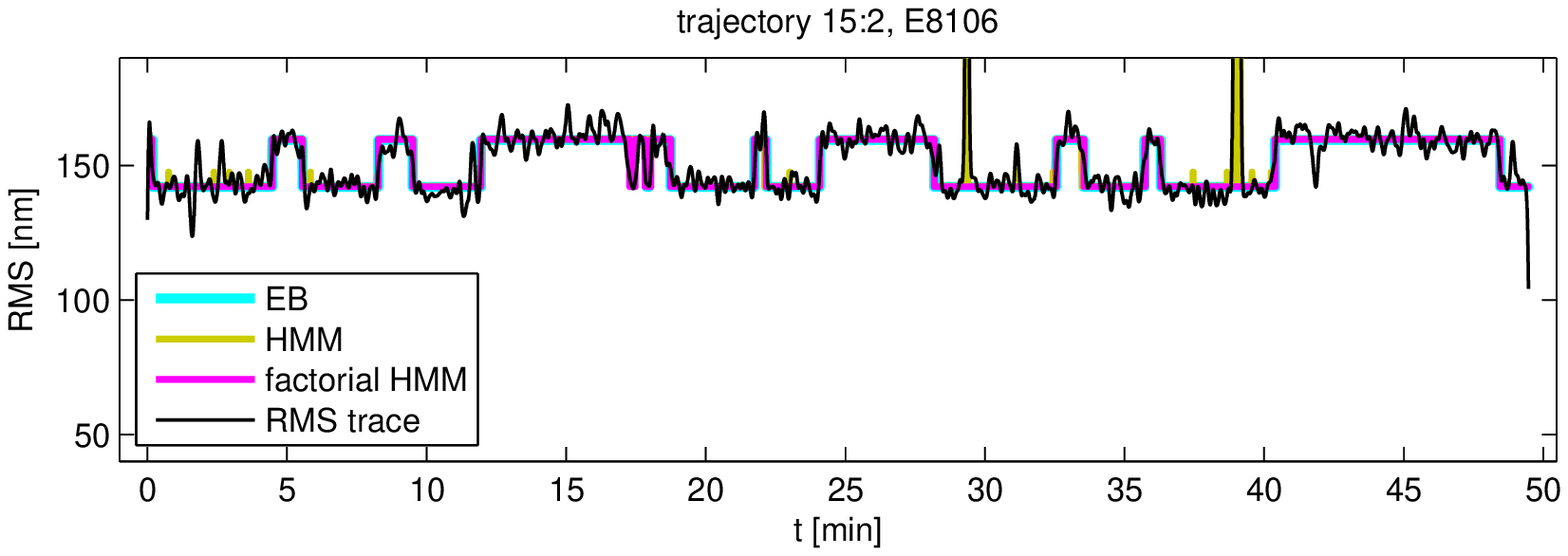}
  \caption{An example of a two-state trajectory of equal length to
    that of Fig.~\ref{fig:ExTrj1}.  Note that there are several short,
    ambiguous excursions of the RMS trace (for example, to a value
    well below that of the looped state around 2 minutes, and to a
    value similar to the looped state around 42 minutes) that would be
    difficult to objectively classify by hand, highlighting one of the
    advantages of the vbTPM approach. The third state is left
    unoccupied by the EB algorithm as well, further confirming the
    2-state classification.}\label{fig:ExTrj2}\end{figure*}

\begin{figure*}
  \includegraphics{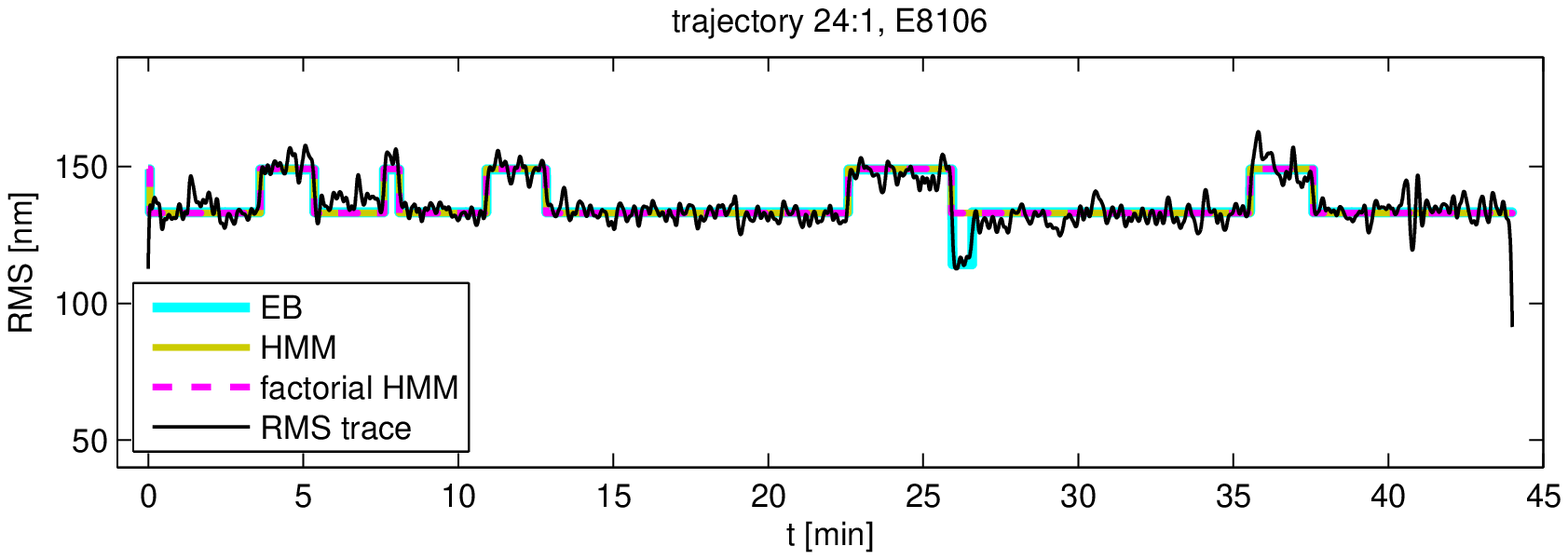}
  \caption{A misclassification of a three-state trajectory as a
    two-state trajectory by the VB algorithm.  The missed third state
    is only visited briefly, around 26 minutes, but recovered by the
    EB algorithm.  Note that here, there were no spurious states, so
    the HMM and factorial HMM overlap
    completely.}\label{fig:ExTrj5}\end{figure*}

%\begin{figure*}
%  \includegraphics{SI_figures/E8106_17_1_segmentation.eps}
%  \caption{}\end{figure*}

%\begin{figure*}
%  \includegraphics{SI_figures/E8106_17_2.eps}
%  \caption{}\end{figure*}

%\begin{figure*}
%  \includegraphics{SI_figures/E8106_33_1.eps}
%  \caption{}\end{figure*}

%\begin{figure*}
%  \includegraphics{SI_figures/E8106_49_1.eps}
%  \caption{}\end{figure*}

\begin{figure*}
  \includegraphics{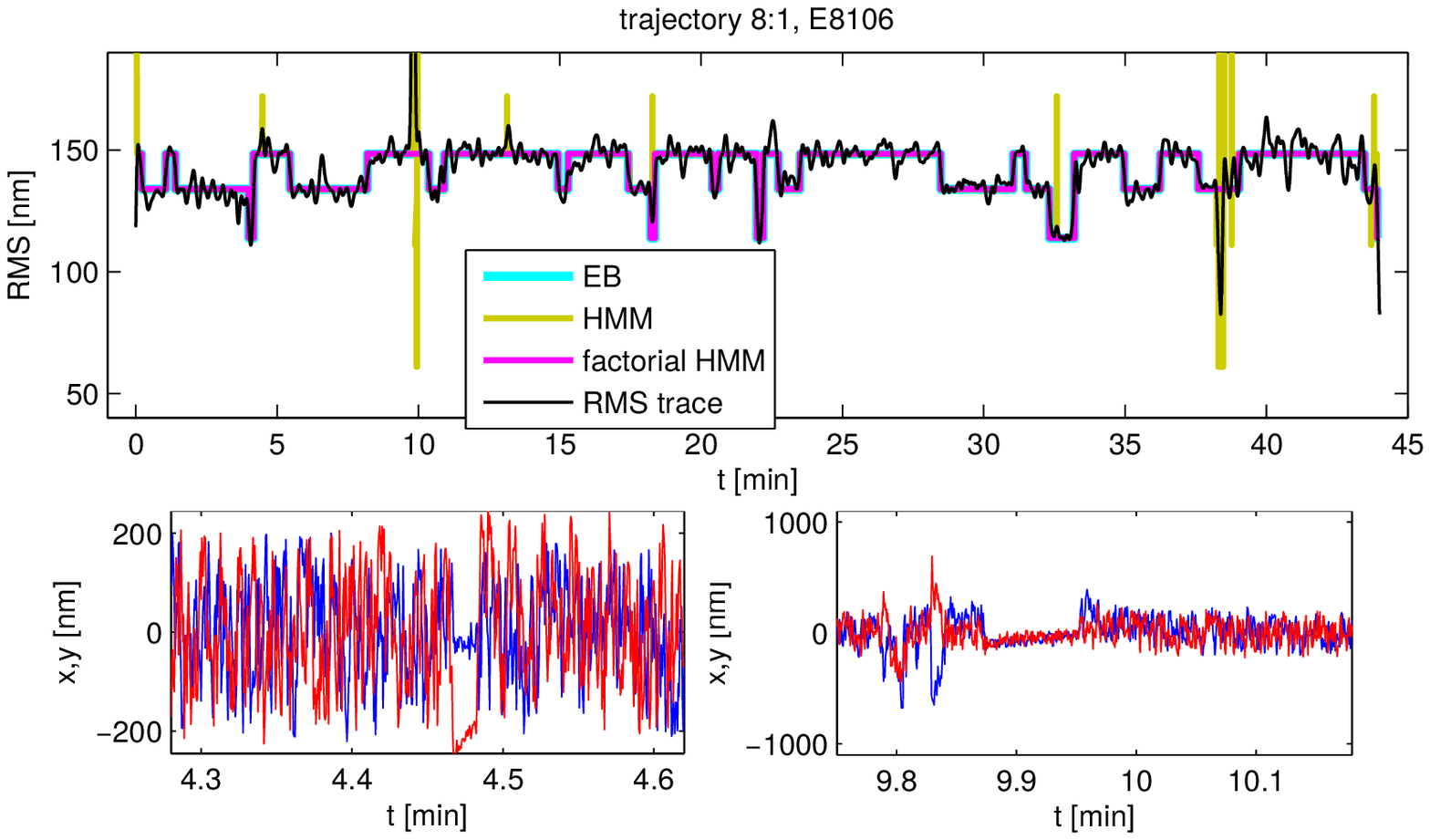}
  \caption{An example of a three-state trajectory with a significant
    number of spurious states, which the factorial HMM successfully
    ignores. The $x(t),y(t)$ positions of some of these spurious
    events are shown in the panels below the main trace, and are
      probably caused by the bead transiently sticking to the surface.
      The slow drift towards the origin during these events are caused
      by the drift-correction filter.}\label{fig:ExTrj3}\end{figure*}

\begin{figure*}
  \includegraphics{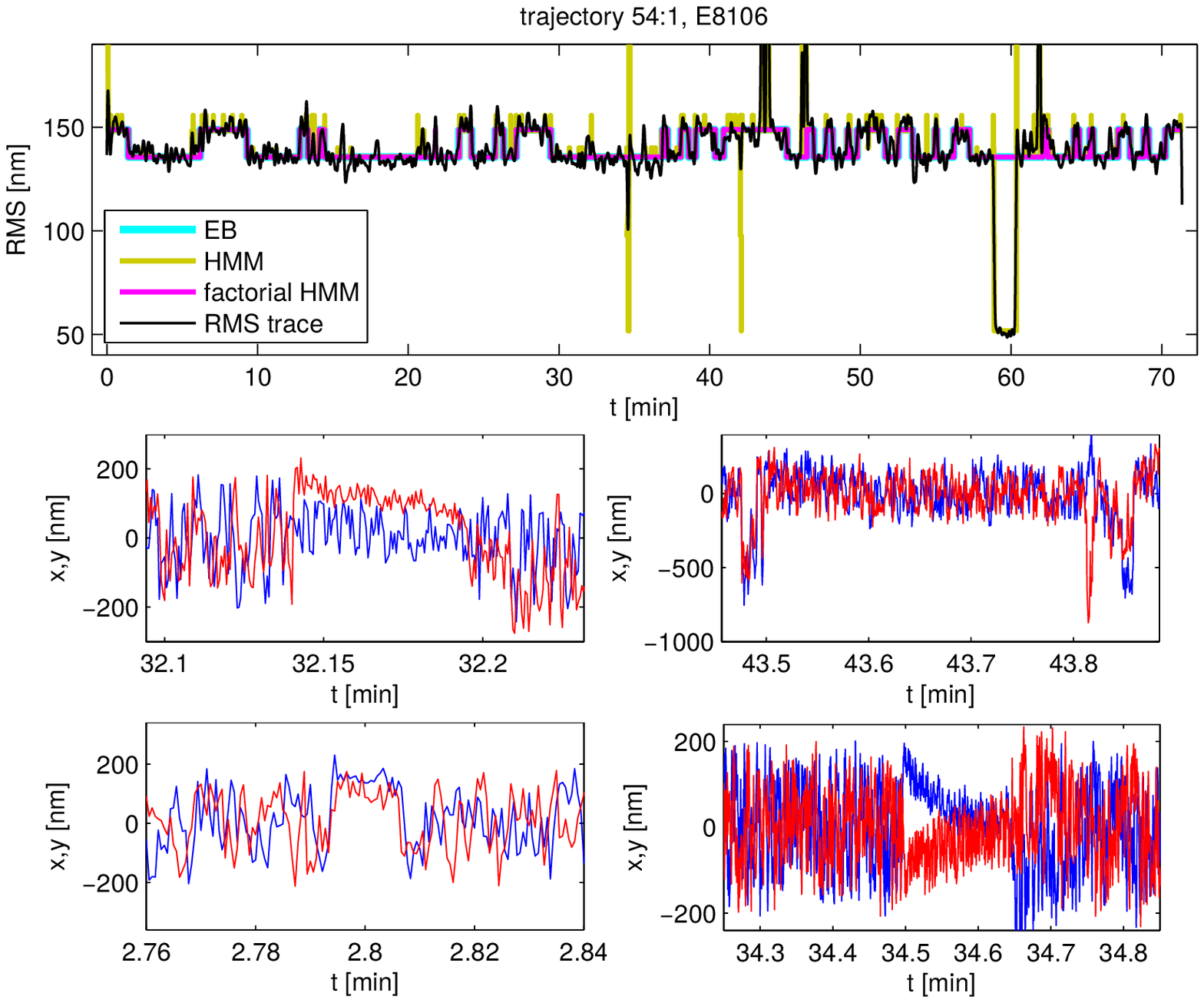}
  \caption{An example of a two-state trajectory with a significant number of spurious states, which the factorial HMM successfully ignores.}\label{fig:ExTrj4}\end{figure*}

%\begin{figure*}
%  \includegraphics{SI_figures/E8106_7_2_segmentation.eps}
%  \caption{}\end{figure*}

\end{document}